\documentclass[prd,twocolumn,superscriptaddress,floatfix,amsmath,amssymb,amsfonts,longbibliography,nofootinbib]{revtex4-1}

\usepackage{float} 

\usepackage[normalem]{ulem}
\usepackage[english]{babel}
\usepackage{graphicx}
\usepackage{dcolumn}
\usepackage{bm}
\usepackage{blindtext}
\usepackage{verbatim}
\usepackage{relsize}
\usepackage{mathrsfs}
\usepackage{musicography}
\usepackage{amsmath}
\usepackage{blindtext}
\usepackage{cancel}
\usepackage{physics}
\usepackage{epstopdf}
\usepackage{mathtools}
\usepackage{tensor}
\usepackage{color}
\usepackage[usenames,dvipsnames]{pstricks}
\usepackage{epsfig}
\usepackage{pst-grad} 
\usepackage{pst-plot} 
\usepackage{hyperref}
\usepackage{verbatim}
\usepackage{slashed}

\newcommand{\mf}{\mathsf}

\newcommand{\ii}{\mathrm{i}}

\newcommand{\tbbb}[1]{\textcolor{black}{#1}}
\newcommand{\tbb}[1]{\textcolor{black}{#1}}
\newcommand{\tb}[1]{\textcolor{black}{#1}}
\newcommand{\trr}[1]{\textcolor{black}{#1}}

\renewcommand{\L}{\mathcal{L}}

\allowdisplaybreaks[1] 

\begin{document}

\title{Harvesting {\color{black} entanglement} from complex scalar and fermionic fields \\with linearly coupled particle detectors}

\author{T. Rick Perche}
\email{trickperche@perimeterinstitute.ca}
\affiliation{Perimeter Institute for Theoretical Physics, Waterloo, Ontario, N2L 2Y5, Canada}
\affiliation{Department of Applied Mathematics, University of Waterloo, Waterloo, Ontario, N2L 3G1, Canada}
\affiliation{Institute for Quantum Computing, University of Waterloo, Waterloo, Ontario, N2L 3G1, Canada}

\author{Caroline Lima}
\email{clima@perimeterinstitute.ca}
\affiliation{Perimeter Institute for Theoretical Physics, Waterloo, Ontario, N2L 2Y5, Canada}
\affiliation{Department of Physics, University of Waterloo, Waterloo, Ontario, N2L 3G1, Canada}

\author{Eduardo Mart\'{i}n-Mart\'{i}nez}
\email{emartinmartinez@uwaterloo.ca}

\affiliation{Perimeter Institute for Theoretical Physics, Waterloo, Ontario, N2L 2Y5, Canada}
\affiliation{Department of Applied Mathematics, University of Waterloo, Waterloo, Ontario, N2L 3G1, Canada}
\affiliation{Institute for Quantum Computing, University of Waterloo, Waterloo, Ontario, N2L 3G1, Canada}

\begin{abstract}
    We explore entanglement harvesting with particle detectors that couple linearly to non-Hermitian fields. Specifically, we analyze the case of particle detectors coupled to a complex scalar quantum field and to a spin 1/2 fermionic field.  \trr{We find that the complex scalar model can be a good approximation for the fermionic model in the protocol of entanglement harvesting when the mass of the field is \tbb{sufficiently large} compared to the inverse interaction time.} \tb{Moreover, we show that taking advantage of the $U(1)$ degree of freedom of a complex detector it is possible to increase the harvested negativity by up to two orders of magnitude \tbb{when compared to the case of a real detector. }}

\end{abstract}

\maketitle

\section{Introduction}

Quantum field theory (QFT) is one of the most successful frameworks of theoretical physics. Among its achievements are the foundations for the standard model and its many applications for the description of numerous other areas of physics ranging from condensed matter to quantum optics. However, our understanding of QFT is still far from complete, and there are many problems still to be fully understood. \tbbb{A relevant example of these kinds of
problems is the lack of a consistent formulation of a measurement
framework in QFT~\cite{formalMeasurements,sorkin,fewster1,fewster2,fewster3,chicken}}. Fundamental studies of {QFT} \tb{both in flat and curved spacetimes} often focus on the properties of the vacuum states. {Indeed}, due to the fact that the short distance behaviour of a QFT is  determined by its vacuum state~\cite{fullingHadamard,fullingHadamard2,kayWald,fewsterNecessityHadamard}, it may be argued that exploring the properties of the vacuum is essential in order to acquire a deeper understanding of QFT.

One of the most remarkable features of the vacuum state of a {quantum field} is the presence of entanglement between spacelike separated regions~\cite{vacuumEntanglement,vacuumBell,Higuchi}. However, quantifying entanglement in {QFTs}, or even defining a local notion of vacuum state is a non-trivial task. A possible way to approach this problem is to quantify the amount of entanglement that can be extracted by local probes that couple to the field at different spacetime regions. In fact, the setup in which two localized probes couple to the vacuum of a quantum field theory {to} spacelike separated regions has become known as entanglement harvesting~\cite{Pozas-Kerstjens:2015,Pozas2016,Petar,Cong2019,foo}.

The first work to study an entanglement harvesting protocol dates back from the 90's, by Valentini \cite{Valentini1991}. This concept was revisited by Reznik et. al. \cite{Reznik2003,Reznik1}, where a pair of pointlike Unruh-DeWitt (UDW) detectors have been used to extract entanglement from the vacuum of a real scalar field. The simplified UDW detector models used by Reznik consisted of two-level systems that couple locally to the quantum field. Multiple different scenarios of entanglement harvesting have been considered during the last decades, including more general couplings~\cite{Sachs1,sachs2018entanglement}, quantum fields~\cite{Pozas2016} and in curved spacetimes~\cite{Nick,Cosmo,Kukita2017,Henderson2019,ericksonBH}.

In this work we study the entanglement harvesting protocol for non-Hermitian quantum fields. We compare the usual scalar field protocols with the case of a complex scalar field and a fermionic field when the detectors-field coupling is linear. Previous literature focused on quadratic couplings~\cite{Sachs1,sachs2018entanglement} that are arguably more difficult to relate to physical processes. In contrast, particle detector models that couple linearly to non-Hermitian fields have recently been linked to physical processes~\cite{neutrinos,antiparticles}. {For instance,} the fermionic particle detectors can be linked to the emission and absorption of neutrinos by nucleons~\cite{neutrinos}. Furthermore, besides computational simplicity and the ability to relate the models to physical processes, there is another advantage of the linear coupling compared to the quadratic one. Namely, the quadratic coupling does not distinguish the particle and the anti-particle sectors of the field, therefore the model cannot capture any effect that depends on the {QFT's} particle versus anti-particle content~\cite{antiparticles}.

We develop the entanglement harvesting formalism for the complex and fermionic fields in curved spacetimes, and then consider examples in flat spacetime, comparing the results with the well-known real scalar field case. We study the role that the particle and antiparticle sectors play in entanglement harvesting and how it relates to the physical process modelled by the detector. Moreover, we highlight what insights from the real scalar case carry through to the linear complex scalar and fermionic case, and which ones do not. 


Our work is organized as follows. In Section \ref{sec:real} we review the standard UDW model, i.e., a two-level system linearly coupled to a real scalar quantum field. We also review the protocol of entanglement harvesting and explicitly study the example where the detectors are initially in the ground and excited states. In Section \ref{sec:anti} we review the linear UDW detectors that couple to a complex scalar field and {in Section \ref{sec:antiHarvest}} we analyze the entanglement harvesting protocol with this model.  In Section \ref{sec:fermion} we review the fermionic linear particle detector model and {in Section \ref{sec:fermiHavest} we} study entanglement harvesting {using this model}. Our conclusions can be found in Section \ref{sec:conclusion}.

\section{The UDW Model and the Entanglement Harvesting setup}\label{sec:real}
    
    In this section we will briefly review the well-known UDW particle detector model, and how it is used in entanglement harvesting scenarios. \textcolor{black}{We then focus in the less commonly explored case where detectors in different states (\tb{excited-ground}) couple to the quantum field.}
    
    \subsection{The UDW Model}\label{sub:UDW}
    
    The {Unruh-DeWitt} detector model in its simplest form~\cite{Unruh1976,DeWitt} consists of a localized two-level quantum system linearly coupled to a free real scalar quantum field. This particle detector model {has} been used extensively in QFT in curved spacetimes and Relativistic Quantum Information in a plethora of scenarios ranging from quantum communication to the study of the Unruh and Hawking effects~(see, among others, \cite{takagi,HawkingRadiation,Unruh-Wald,matsasUnruh,waitUnruh,fewsterUnruh,edunruh,unruhEffectNoThermal,unruhSlow,mine}). {This model has also} been proven to capture the fundamental features of the light-matter interaction when exchange of angular momentum is not relevant~\cite{eduardoOld,Pozas2016,richard}.  In particular for our purposes, particle detectors have been used to study the entanglement structure of quantum fields through the protocol known as \textit{entanglement harvesting}~\cite{Pozas-Kerstjens:2015,Pozas2016,Petar,Cong2019,foo}.
    
    In order to introduce the UDW model, we consider a $D = n+1$ dimensional globally hyperbolic curved spacetime {\color{black}$M$} with a Lorentzian metric $\mathsf{g}$. In this spacetime we introduce a real scalar quantum field $\hat{\phi}(\mf x)$. We assume that the scalar field can be canonically quantized in terms of a normalized basis of solutions of the Klein-Gordon equation, $\{u_{\bm k}(\mf x),u_{\bm k}^*(\mf x)\}$, according to
    \begin{equation}\label{eq:scalarField}
        \hat{\phi}(\mf x) = \int \dd^n \bm k \left(u_{\bm k}(\mf x) \hat{a}_{\bm k}+u_{\bm k}^*(\mf x) \hat{a}^\dagger_{\bm k}\right),
    \end{equation}
    where the creation and annihilation operators, $\hat{a}^\dagger_{\bm k}$ and $\hat{a}_{\bm k}$, satisfy the bosonic canonical commutation relations
    \begin{equation}
        \comm{\hat{a}^{\vphantom{\dagger}}_{\bm k}}{\hat{a}^\dagger_{\bm k'}} = \delta^{(3)}(\bm k - \bm k').
    \end{equation}
    
    We then consider a particle detector moving in a timelike trajectory $\mf z(\tau)$ in $\tbb{M}$, where $\tau$ is the proper time parameter of the curve.
    The free Hamiltonian of the detector that generates time evolution with respect to its proper time is given by
        \begin{equation}\label{eq:detectorH}
            \hat{H}_\text{d} = \Omega\hat{\sigma}^+\hat{\sigma}^-,
        \end{equation}
    where $\Omega$ is the energy gap between the ground and excited states ($\ket{g}$ and $\ket{e}$) of the detector and $\hat{\sigma}^\pm$ are the ladder operators. Namely,  $\hat{\sigma}^+ = \ket{e}\!\!\bra{g}$ and $\hat{\sigma}^- = \ket{g}\!\!\bra{e}$.

    The coupling between the field and the detector is given by the following interaction Hamiltonian weight\footnote{The Hamiltonian weight $\hat{h}_I(\mf x)$ is related to the Hamiltonian density $\mathfrak{h}_I(\mf x)$ by $\mathfrak{h}_I(\mf x) = \hat{h}_I(\mf x) \sqrt{-g}$.} in the interaction picture~\cite{us,us2}:
    \begin{equation}\label{eq:hIreal}
        \hat{h}_I(\mf x) = \lambda \Lambda(\mf x) \hat{\mu}(\tau) \hat{\phi}(\mf x),
    \end{equation}
    where $\lambda$ is the coupling strength, $\Lambda(\mf x)$ is the spacetime smearing function, responsible for controlling both the spatio-temporal profile of the interaction and \mbox{$\hat{\mu}(\tau) = e^{\mathrm{i}\Omega\tau}\hat{\sigma}^+ + e^{-\mathrm{i}\Omega\tau}\hat{\sigma}^-$} is the detector's monopole moment in the interaction picture. The associated time evolution operator is then given by\footnote{
    It is usual to write the integral in Eq.~\eqref{UI} in terms of an integral in time of a Hamiltonian associated with a given foliation $\Sigma_t$ using 
    \begin{equation*}
        \int \dd V = \int \dd t \int_{\Sigma_t} \dd^n \bm x \sqrt{-g},
    \end{equation*}
    where $t$ is a time parameter that parametrizes a foliation by spacelike surfaces $\Sigma_t$ for which we use an arbitrary spatial coordinate $\bm x$ (for more details see~\cite{us,us2}). The Hamiltonian associated to this foliation is then given by \begin{equation*}
    \hat{H}_I(t) = \int_{\Sigma_t} \dd^n \bm x\: \sqrt{-g}\,\hat{h}_I(\mf x). 
    \end{equation*}
    }
    \begin{align}\label{UI}
        \hat{U}_I &= \mathcal{T}_\tau \exp \left(-\mathrm{i}\int \dd V \hat{h}_I(\mf x)\right),
    \end{align}
    where $\dd V$ is the invariant spacetime volume element and $\mathcal{T}_\tau$ denotes the time ordering operation with respect to the proper time parameter $\tau$. It is important to notice that for smeared particle detectors, the time ordering operation, in principle, can depend on the time parameter chosen. However, in~\cite{us2} it was shown that under the right conditions, any choice of time ordering is equivalent up to leading order in the coupling constant.

    It is common to work perturbatively with the Dyson expansion for the time evolution operator $\hat{U}_I$, so that, to second order, it is given by
    \begin{equation}
        \hat{U}_I = \openone + \hat{U}_I^{(1)} + \hat{U}_I^{(2)} + \mathcal{O}(\lambda^3), 
    \end{equation}
    with   
    \begin{equation}
        \begin{aligned}
            \hat{U}_I^{(1)} &= -\mathrm{i}\int \dd V \hat{h}_I(\mf x),\\
            \hat{U}_I^{(2)} &= -\int \dd V\dd V' \hat{h}_I(\mf x) \hat{h}_I(\mf x')\theta(\tau-\tau'),
        \end{aligned}
    \end{equation}
    where $\theta(\tau)$ denotes the Heaviside theta function that arises from the time ordering operation.\footnote{Here $\tau$ denotes the time-like Fermi normal coordinates associated with the trajectory $\mf z(\tau)$. This is important for spatially smeared detectors and in fact $\tau = \tau(\mf x)$ extends the notion of its proper time to a local region around the curve. More details can be found in~\cite{us,us2}.}
   
    We assume the detector to start in a state $\hat{\rho}_{\text{d},0}$ and to be completely uncorrelated with the field state, $\hat{\rho}_{\phi}$. That is, we consider the initial state of the detector-field system to be the density operator $\hat{\rho}_0 = \hat{\rho}_{\text{d}, 0} \otimes \hat{\rho}_{\phi}$. After time evolution the final state of the full system will be given by
    \begin{equation}
        \label{eqtimeevolution}
        \hat{\rho} = \hat{U}_I\hat{\rho}_0\hat{U}_I^\dagger.
    \end{equation}
    We can obtain the evolved state of the detector by tracing out the field degrees of freedom. Up to second order in $\lambda$, we get
    \begin{equation}\label{rhoD}
        \hat{\rho}_\text{d} = \hat{\rho}_{\text{d}, 0}+\hat{\rho}_{\text{d}}^{(1)}+\hat{\rho}_{\text{d}}^{(2)}+\mathcal{O}(\lambda^3),
    \end{equation}
    where 
    \begin{equation}
        \begin{aligned}
            \label{eqrhoDexpansions}
            \hat{\rho}_{\text{d}}^{(1)} &= \tr_{\phi}\left(\hat{U}_I^{(1)}\hat{\rho}_0+\hat{\rho}_0 \hat{U}_I^{(1)\dagger}\right),\\
            \hat{\rho}_{\text{d}}^{(2)} &= \tr_{\phi}\left(\hat{U}_I^{(2)}\hat{\rho}_0+\hat{U}_I^{(1)}\hat{\rho}_0 \hat{U}_I^{(1)\dagger}+\hat{\rho}_0 \hat{U}_I^{(2)\dagger}\right),
        \end{aligned}
    \end{equation}
    where $\tr_\phi$ denotes the trace over the field degrees of freedom. 
    
    The excitation probability {for} a detector initially in the ground state and a field in an arbitrary state $\hat{\rho}_{\phi}$ is given by
    \begin{align}
        p_{g\rightarrow e} =& \tr(\hat{\rho}_{\text{d}} \ket{e}\!\!\bra{e})\\
        =& \lambda^2\int \dd V \dd V' \Lambda(\mf x) \Lambda(\mf x') e^{\mathrm{i} \Omega(\tau-\tau')}\langle \hat{\phi}(\mf x') \hat{\phi}(\mf x)\rangle_{\hat{\rho}_{\phi}} \nonumber\\ &\tb{+\mathcal{O}(\lambda^3)}.\nonumber
    \end{align}
    The excitation probability of the detector allows for the study of numerous features of quantum field theories, such as particle production by numerous effects, including Hawking radiation, detector acceleration and external effects in the field.
    
    Further, if the field starts in a Gaussian state with zero mean, the time evolution of the detector coupled to the field will depend on the field only through its Wightman function. For example, if the field starts in the vacuum state $\ket{0}$ defined by  $\hat{a}_{\bm k}\ket{0} = 0$ for all $\bm k$, where $\hat{a}_{\bm k}$ are the annihilation operators defined in the mode expansion~\eqref{eq:scalarField},  we get
    \begin{equation}\label{wPhi}
        \bra{0}\!\hat{\phi}(\mf x') \hat{\phi}(\mf x)\!\ket{0} = \int \dd^n \bm k\, u_{\bm k}(\mf x')u_{\bm k}^*(\mf x).
    \end{equation}
    These considerations will become useful when we particularize our studies to vacuum entanglement harvesting.
    

    
\subsection{{Ground-Ground Entanglement Harvesting Protocol}}\label{sub:UDWharvest}

    Now that we have introduced the interaction between one detector and the field, we will investigate the entanglement harvesting protocol~\cite{Pozas-Kerstjens:2015,Pozas2016,Petar,Cong2019,foo}. In this protocol, instead of one, we must couple (at least) two particle detectors to the field. Our main interest is to analyze the harvesting of quantum correlations between spacelike separated detectors.
    
    Let us consider two detectors undergoing timelike trajectories $\mf z_i (\tau_i)$ for $i=1,2$. Each of the two detectors interact with the field as in Eq. \eqref{eq:hIreal}. That is, in the interaction picture, the interaction Hamiltonian weight of each detector $i$ is
        \begin{equation}
        \hat{h}_{I, i}(\mf x) = \lambda_i \Lambda_i(\mf x) \hat{\mu}_i(\tau_i) \hat{\phi}(\mf x),
    \end{equation}
    where $\lambda_i$ are the detectors respective coupling constants.
    
    The full interaction Hamiltonian weight will then be given by
    \begin{equation}\label{hIfull}
        \hat{h}_{I}(\mf x) =  \hat{h}_{I,1}(\mf x)+\hat{h}_{I,2}(\mf x).
    \end{equation}
    We assume that the field and the detectors have no correlations prior to the interaction. That is, the detectors-field system density operator will be given by \mbox{$\hat{\rho}_0 = \hat{\rho}_{\text{d}, 0} \otimes \hat{\rho}_{\phi}$}, where $\hat{\rho}_{\text{d}, 0} = \hat{\rho}_{\text{d}_1, 0} \otimes \hat{\rho}_{\text{d}_2, 0}$. Here $\text{d}$ labels the two-detectors subsystem and d$_i$ labels the $i$-th detector. $\hat{\rho}_0$ will then evolve according to Eq. \eqref{eqtimeevolution}{\color{black}. As previously mentioned, the time evolution operator will depend on a notion of time ordering. Unlike the case where we consider one detector, in this general scenario with two detectors undergoing arbitrary motion, it is even more ambiguous what is the best notion  of time ordering since their proper times may be radically different. However, as we discussed above, provided that the detectors start in either the ground or excited state, any notion of time ordering will yield the same result to leading order as shown in \cite{us2}. Given that there is no reason to pick time ordering with respect to one detector or the other and that to leading order the choice of time parameter for the ordering is irrelevant, we simply denote the time coordinate used to prescribe the time ordering by $t$, so that we can write
    \begin{equation}\label{itstnow}
        \hat{U}_I = \mathcal{T}_t \exp\left(-\text{i} \int \dd V \hat{h}_I(\mf x)\right),
    \end{equation}
    where $\mathcal{T}_t$ denotes the time ordering operator with respect to a parameter $t$, and $\hat{h}_I(\mf x)$ is given by Eq. \eqref{hIfull}.
    
    To leading order in $\lambda$, Eqs. \eqref{rhoD} and \eqref{eqrhoDexpansions} hold with $\hat{U}_I$ as in Eq. \eqref{itstnow}.} We then obtain the final state of the two-detectors subsystem by tracing out the field state. If both detectors are initially in their ground state (that is $\hat{\rho}_{\text{d},0} = \ket{g_1g_2}\!\bra{g_1g_2}$), their density operator  in the $\{\ket{g_1g_2}, \ket{g_1e_2}, \ket{e_1g_2}, \ket{e_1e_2}\}$ basis will be
    \begin{equation}\label{eq:rhoDscalar}
        \hat{\rho}_{\text{d}} = \begin{pmatrix}
            1 - \L_{11}-\L_{22} & \mathrm{i}\mathcal{E}_2\color{black} & \mathrm{i}\mathcal{E}_1\color{black} & \mathcal{M}^*\\
             -\mathrm{i}\mathcal{E}_2^* & \L_{22} & \L_{21} & 0\\
            -\mathrm{i}\mathcal{E}_1^* & \L_{12} & \L_{11} & 0\\
            \mathcal{M} & 0 & 0 & 0 
        \end{pmatrix} + \mathcal{O}(\lambda^3),
    \end{equation}
    where
\begin{align}
        \L_{ij} &=\!\lambda_i\lambda_j\!\!\int \dd V \dd V'\Lambda_i(\mf x)\Lambda_j(\mf x')e^{\mathrm{i}\Omega_i \tau_i -\mathrm{i}\Omega_j \tau_j'}\langle\hat{\phi}(\mf x')\hat{\phi}(\mf x) \rangle_{{\rho}_{\phi}},\nonumber\\ 
        \mathcal{M} &= - \lambda_1\lambda_2 \int \dd V\dd V'\theta(t - t') \nonumber\\
        &\:\:\:\:\:\:\:\:\Big(\Lambda_1(\mf x)\Lambda_2(\mf x')e^{\mathrm{i}\Omega_1 \tau_1 + \mathrm{i}\Omega_2 \tau_2'}\langle\hat{\phi}(\mf x)\hat{\phi}(\mf x')\rangle_{\rho_{\phi}} \nonumber\\
        &\:\:\:\:\:\:\:\:\:\:\:\:\:\:\:\:+\Lambda_2(\mf x)\Lambda_1(\mf x')e^{\mathrm{i}\Omega_2 \tau_2 +\mathrm{i}\Omega_1 \tau_1'}\langle\hat{\phi}(\mf x)\hat{\phi}(\mf x')\rangle_{\rho_{\phi}}\Big), \nonumber\\
       \mathcal{E}_i &    = \lambda_i \int \dd V \Lambda_i(\mf x) e^{-\mathrm{i} \Omega_i \tau_i} \langle\hat{\phi}(\mf x)\rangle_{\rho_{\phi}}.\label{expReal}
    \end{align}
    {\color{black} Notice that the only term of the above that depends on the time ordering operation is $\mathcal{M}$, which can be seen from the dependence on $\theta(t-t')$. The details of these computations can be found in Appendix \ref{app:gg}.}

    There are several common entanglement measures that are typically used in entanglement harvesting. The two most popular are concurrence \tbb{$\mathcal{C}$}~\cite{WootersConcurence} and negativity \tbb{$\mathcal{N}$}~\cite{WootersAlone,VidalNegativity}. \tb{In general, they both have their advantages and disadvantages. The main advantage of negativity versus concurrence is that negativity can be used for higher dimensional bipartite states, and not only qubits. This allows for a more direct comparison between two-level detectors and harmonic oscillator detectors~\cite{HOdetectorsUnruh,HOdetectorsBeiLok,HODetectorsAcc}\footnote{\tb{Note that the negativity is a proper measure of entanglement for states of two qubits and Gaussian states of two harmonic oscillators. Although it is not true in general, a non-zero negativity is necessary and sufficient for the state to be entangled in these cases.}}. On the other hand, the concurrence (which is only defined for two two-dimensional systems) can be more easily related to the entanglement of formation, which is a very intuitive measure of entanglement. \tbb{Furthermore, the negativity is always bounded by twice the concurrence. What is more, for the examples analyzed in this paper, we will always have that $\mathcal{E}_1 = \mathcal{E}_2 = 0$, and in these cases the concurrence and negativity are equivalent (in fact $\mathcal{N}=2\mathcal{C}$) so that the entanglement of formation is also a monotonically increasing function of the negativity~\cite{topology}}. For the reasons stated above, we will use negativity \tbb{to quantify the entanglement acquired by the detectors}, which}, unlike concurrence, can be readily used both for qubit detectors and higher dimensional ones.
    
    The negativity of a bipartite density operator is defined as the  sum of the absolute value of the negative eigenvalues of its partial transpose. In the case of the density operator in Eq. \eqref{eq:rhoDscalar}, the negativity is given by $\mathcal{N}=\mathcal{N}^{(2)}+\mathcal{O}(\lambda^3)$, where $\mathcal{N}^{(2)}$ is defined as the maximum between $0$ and
    \begin{widetext}
    \begin{equation}\label{UDWN}
        \!\frac{1}{2}\!\!\left(\!\!\sqrt{\!(\L_{11} -\L_{22})^2\!+\!(|\mathcal{E}_1|^2+|\mathcal{E}_2|^2){}^{{}^2}\!\!+\!2(|\mathcal{E}_1|^2\!-\!|\mathcal{E}_2|^2)(\mathcal{L}_{11}\! - \!\mathcal{L}_{22}) \!-\! 8\Re(\mathcal{E}_1\mathcal{E}_2^*\mathcal{M})\!+\!\abs{4\mathcal{M}}^2} \!-\!\mathcal{L}_{11}\! -\! \mathcal{L}_{22}\!+\!|\mathcal{E}_1|^2\!+\!|\mathcal{E}_2|^2\!\right)\!.
    \end{equation}
    \end{widetext}
For setups where $\mathcal{L}_{11} = \mathcal{L}_{22} = \mathcal{L}$ and $\mathcal{E}_1 = \mathcal{E}_2 = \mathcal{E}$ {(e.g., identical detectors, identically coupled, interacting in flat spacetime)}, Eq. \eqref{UDWN} simplifies to
    \begin{equation}
        \mathcal{N}^{(2)} = \max\left(0, |\mathcal{M} - |\mathcal{E}|^2|+|\mathcal{E}|^2- \mathcal{L}\right).
    \end{equation}

\textcolor{black}{Entanglement harvesting for detectors initially in their ground state has been extensively studied in the literature (see, among many others, \cite{Pozas-Kerstjens:2015,Pozas2016,Petar,Cong2019,foo,Nick,Cosmo,Kukita2017,Henderson2019,ericksonBH}). Since many examples of the harvesting protocol with this initial state can be found in the literature, we will not show any particular example in this subsection.}

\subsection{{\tb{Excited-Ground} Entanglement Harvesting Protocol}\label{sub:UDWharvestge}}

We now consider the case (less frequently studied in the literature) where one detector starts in the excited state
, while the other detector starts in the ground state
. In this setup we can write the initial state of the detectors-field system, $\hat{\rho}_0$  as
    \begin{equation}\label{eq:geinitialscalar}
        \hat{\rho}_0 = \ket{e_1}\!\!\bra{e_1}\otimes\ket{g_2}\!\!\bra{g_2}\otimes \hat{\rho}_{\phi} = \hat{\rho}_{\text{d}, 0}\otimes \hat{\rho}_{\phi}.
    \end{equation}
    
    Performing the computations in the same fashion as we did in the ground-ground scenario, it is possible to obtain the time-evolved density operator associated with the two detectors. The full detail of {this} computation can be found in Appendix \ref{app:ge}. The time-evolved density matrix for the two detectors after coupling to the scalar field is
    \begin{equation}
        \hat{\rho}_{\text{d}} = \begin{pmatrix}
            \tb{\mathcal{L}_{11}'} & 0 & -\mathrm{i}\mathcal{E}_1 & \tb{\mathcal{L}_{12}'}\\
            0 & 0 & \mathcal{M}' & 0 \\
            \mathrm{i}\mathcal{E}_1^* & \mathcal{M}'^* & 1 - \tb{\mathcal{L}_{11}'} - \mathcal{L}_{22} & \mathrm{i}\mathcal{E}_2\\
            \tb{\mathcal{L}_{12}'^*} & 0 & -\mathrm{i}\mathcal{E}_2^* & \mathcal{L}_{22}        
        \end{pmatrix}+ \mathcal{O}(\lambda^3),
    \end{equation}
    where $\mathcal{L}_{ij}$ and $\mathcal{E}_i$ are given by Eq. \eqref{expReal} and the remaining components are explicitly given by
    \begin{align}
        \tb{\mathcal{L}_{12}'}\! =& \tb{\lambda_1\lambda_2}\!\!\int\!\!\dd V\dd V'\!\Lambda_1(\mf x)\Lambda_2(\mf x')e^{\!-\mathrm{i}(\Omega_1 \tau_1 +\Omega_2 \tau_2')} \!\langle\hat{\phi}(\mf x) \hat{\phi}(\mf x')\rangle_{\!\hat{\rho}_{\!\phi}}\nonumber,\\
        \mathcal{M}'{} =& \tb{\lambda_1 \lambda_2}  \int \dd V \dd V' \theta(t-t')\label{Nij} \\&\Big(\Lambda_1(\mf x)\Lambda_2(\mf x')e^{\mathrm{i}\Omega_2 \tau_2'-\mathrm{i}\Omega_1 \tau_1}\!\langle \hat{\phi}(\mf x)\hat{\phi}(\mf x') \rangle_{\rho_{\phi}}\nonumber\\
        &
        \:\:\:\:\:\:\:\:\:\:\:\:\:+\Lambda_2(\mf x)\Lambda_1(\mf x')e^{\mathrm{i}\Omega_2 \tau_2 -\mathrm{i}\Omega_1 \tau_1'}\!\langle\hat{\phi}(\mf x)\hat{\phi}(\mf x')\rangle_{\rho_{\phi}} \!\Big)\nonumber,\\
         \tb{\mathcal{L}_{11}'} =& \tb{\lambda_1^2} \int \dd V \dd V'\Lambda_1(\mf x) \Lambda_1(\mf x') e^{-\ii \Omega_1(\tau_1 - \tau_1')}\langle\hat{\phi}(\mf x') \hat{\phi}(\mf x)\rangle_{\rho_{\phi}}.\nonumber
    \end{align}
    When compared to the case of ground-ground initial state, the $\mathcal{M}'{}$ terms play the role of the $\mathcal{M}$ terms and the $\tb{\mathcal{L}_{12}'}$ term play the role of the $\mathcal{L}_{12}$ terms {and $\tb{\mathcal{L}_{11}'}$ represent the deexcitation probability of the first detector}. Notice that $\mathcal{M}'$ has different signs for the phases $\Omega_1\tau_1$ and $\Omega_2\tau_2$, while  $\mathcal{M}$ has the same sign for these phases. This is not surprising since the change $\Omega_i\to-\Omega_i$ swaps the ground and the excited states (gap inversion). We can say that the integrand of $\mathcal{M}$ is `faster rotating', while the integrand of $\mathcal{M}'{}$ is `slower rotating'. The opposite happens with the \tb{$\mathcal{L}_{12}$ terms, which are slower rotating, while the $\tb{\mathcal{L}_{12}'}$ terms from the excited-ground setup are faster rotating.} For numerical purposes, faster rotating terms are harder to integrate when compared to slower rotating terms. In particular, this makes the {calculation} of the harvested entanglement easier when the initial state of the detectors is \tb{excited-ground} \tb{due to the fact that the negativity only depends on $\mathcal{M}'$ and not on $\tb{\mathcal{L}_{12}'}$}. 
    
    The negativity for this choice of initial state is still given by a similar expression to~\eqref{UDWN}. When the field's state is such that $\langle\hat{\phi}(\mf x)\rangle_{\rho_{\phi}} = 0$ {(hence, $\mathcal{E}_1 = \mathcal{E}_2 = 0$)} the negativity is explicitly given by :
    \begin{equation}\label{negrge}
        \mathcal{N}^{(2)} = \max\left(0,\sqrt{{|\mathcal{M}'|}{}^2+\frac{({\tb{\L_{11}'}} -\L_{22})^2}{4} }-\frac{\tb{\L_{11}'} + \mathcal{L}_{22}}{2}\right),
    \end{equation}
    but---related to the fact that $\mathcal{M}'{}$ is slower rotating than $\mathcal{M}$---the negativity will be larger than what would be found in the ground-ground protocol.
    
    \subsection{\tb{Explicit example in flat spacetime}}\label{sub:egex}
    Although the literature is full of examples of ground-ground entanglement harvesting, the \tb{excited-ground} protocol is not so commonly analyzed. For this reason, besides considering the  general case above we will also go into more explicit detail for two comoving inertial detectors in Minkowski spacetime. Specifically, let us consider the mode expansion of the quantum field to be given by
    \begin{equation}
        u_{\bm p}(\mf x) = \frac{1}{(2\pi)^{\frac{n}{2}}} \frac{e^{\mathrm{i} \mf p \cdot \mf x}}{\sqrt{2 \omega_{\bm p}}},
    \end{equation}
    where we assume inertial coordinates with $\mf x = (t,\bm x)$ and we write $\mf p = (\omega_{\bm p},\bm p)$, with $\omega_{\bm p} = \sqrt{\bm p^2 + m^2}$, where $m$ is the field's mass. 
    
    The spacetime smearing functions for the detectors will be prescribed as
    \begin{align}
        \Lambda_1(\mf x) &= \chi(t)F(\bm x),\label{Lambda1}\\
        \Lambda_2(\mf x) &= \chi(t)F(\bm x-\bm L),\label{Lambda2}
    \end{align}
    where we split the spacetime smearing into a switching function $\chi(t)$ and a smearing function $F(\bm x)$. The choices above assume the detectors to be comoving with the $(t,\bm x)$ coordinate system with detector $1$ located at the origin and detector $2$ located at $\bm L$. Without loss of generality, we assume $\bm L = L \bm e_z$, where $\bm e_z$ is the unit vector in the $z$ direction. We also assume $\Omega_1 = \Omega_2 = \Omega$, which makes the detectors identical. \tb{Moreover, we will consider $\lambda_1 = \lambda_2$ in this example.} With these assumptions, we can rewrite the negativity using {$\tb{\mathcal{L}_{11}'} = \L(-\Omega)$ and \mbox{$\L_{22} = \L(\Omega)$}}, where     
    \begin{align}\label{groundexcitedM}
        &\mathcal{L}{(\Omega)} = \frac{\lambda^2}{(2\pi)^n}\!\int \frac{\dd^n \bm p}{2 \omega_{\bm p}}|\tilde{F}(\bm p)|^2 |\tilde{\chi}( \omega_{\bm p}+\Omega)|^2,\\
        &\mathcal{M}' \!=\!\frac{\lambda^2}{(2\pi)^n} \!\int \!\frac{\dd^n \bm p}{2 \omega_{\bm p}} |\tilde{F}(\bm p)|^2 e^{\mathrm{i} \bm p \cdot \bm L}\!\left(Q(\Omega\!-\!\omega_{\bm p})\!+\!Q(\!-\omega_{\bm p}\!-\!\Omega)\right)\!,\label{groundexcitedM2}
    \end{align}
    where $ Q(\omega)$ is defined as
    \begin{equation}\label{Comega}
        Q(\omega) = \int \dd t \dd t' \theta(t-t') e^{\mathrm{i}\omega( t - t')} \chi(t)\chi(t'),
    \end{equation}
    and tilde denotes the Fourier transform, defined as
    \begin{equation}
        \tilde{F}(\bm p)  = \int \dd^n \bm x F(\bm x) e^{\mathrm{i} \bm p \cdot \bm x},\quad
        \tilde{\chi}(\omega) = \int \dd t \chi(t) e^{\mathrm{i} \omega t}.
    \end{equation}
    

     In order to study an explicit example, we prescribe the switching and smearing functions according to
    \begin{align}
        \chi(t) &= \frac{1}{\sqrt{2 \pi}}e^{-\frac{t^2}{2 T^2}},\label{chi}\\
        F(\bm x) &= \frac{1}{(2 \pi \sigma^2)^\frac{n}{2}}e^{-\frac{\bm x^2}{2 \sigma ^2}}.\label{F}
    \end{align}
    The parameters $\sigma$ and $T$ control the size of the interaction region and the duration timescale of the interaction, respectively. \tbb{With the choices of Eqs. \eqref{chi} and \eqref{F} we can consider the detectors to be approximately spacelike separated for $L\geq 5T\:$ \footnote{\tbb{Although the detector size (controlled by $\sigma$) is in general relevant for determining whether the detectors are spacelike separated we will always consider $\sigma \leq  0.1 L$, so that $L\geq 5 T$ ensures effective spacelike separation.}}. We then} have
    \begin{align}
        Q(\omega) &= T^2 e^{-\omega^2 T^2} \left( 1 - \text{Erf}(-\mathrm{i} \omega T)\right),\label{expC}\\
        \tilde{\chi}(\omega) &= T e^{-\frac{ \omega^2T^2}{2}},\quad \quad \tilde{F}(\bm p) = e^{- \frac{\sigma^2 \bm p^2}{2}}.\label{expF}
    \end{align}
    With the results above, it is possible to find analytical expressions for the integrals over the angular variables, so that $\mathcal{L}{(\Omega)}$ and $\mathcal{M}'$ can be written as a single integral over $|\bm p|$. We perform the computations in $n$ dimensions in Appendix \ref{app:angles}. In the particular case of three space dimensions, we have:
    \begin{align}
        \label{resultgroundexcited}
        \mathcal{L}{(\Omega)} = \frac{\lambda^2}{2\pi^2} \int_0^\infty \frac{\dd |\bm  p|}{2 \omega_{\bm p}} &|\bm p|^2 \:|\tilde{F}(|\bm p|)|^2 |\tilde{\chi}( \omega_{\bm p}+\Omega)|^2,\nonumber\\
        \mathcal{M}'
        =\frac{\lambda^2}{{2\pi^2}} \int_0^\infty \frac{\dd |\bm  p|}{2 \omega_{\bm p}} &|\bm p|^2 \:|\tilde{F}(|\bm p|)|^2 \text{sinc}( |\bm p| L)\\ &\trr{\times\Big(Q(\Omega-\omega_{\bm p})+Q(-\omega_{\bm p}-\Omega)\Big)},\nonumber
    \end{align}
    At last, the integral over $|\bm p|$ cannot be solved in terms of elementary functions, so we resort to numerical integration.
    
     \begin{figure}[h]
        \includegraphics[scale=0.36]{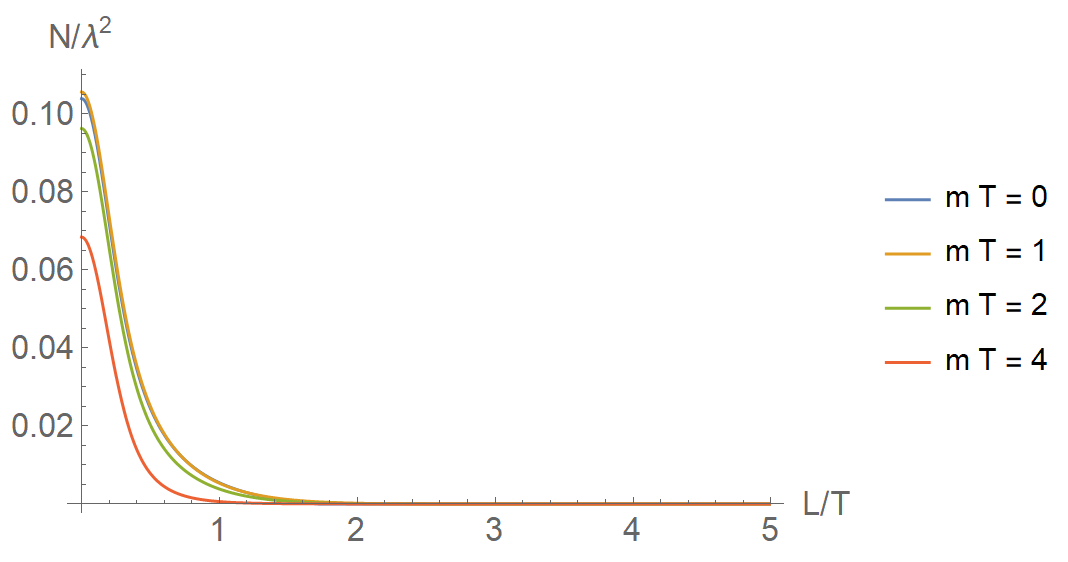}
        \caption{Negativity of the two detector system as a function of the detectors separation for different values of the field mass. We fixed the detectors gap as $\Omega T = 1$ and the detector size as $\sigma = 0.1 T$. }\label{fig:scalar}
    \end{figure}
    \begin{figure}[h]
        \includegraphics[scale=0.36]{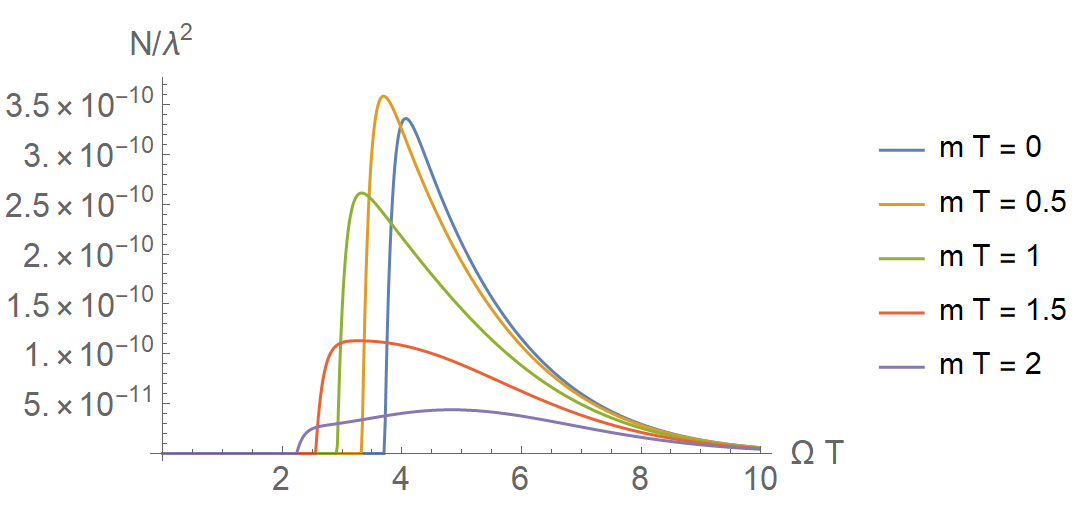}
        \caption{Negativity of the two detector system as a function of the detector gap for different values of the field mass. We fixed the detector separation as $L = 5 T$ and the detector size as $\sigma = 0.2 T$ (this value was chosen to ease the numerical evaluation).}\label{fig:scalar1}
    \end{figure}
    
    \begin{figure}[h]
        \includegraphics[scale=0.36]{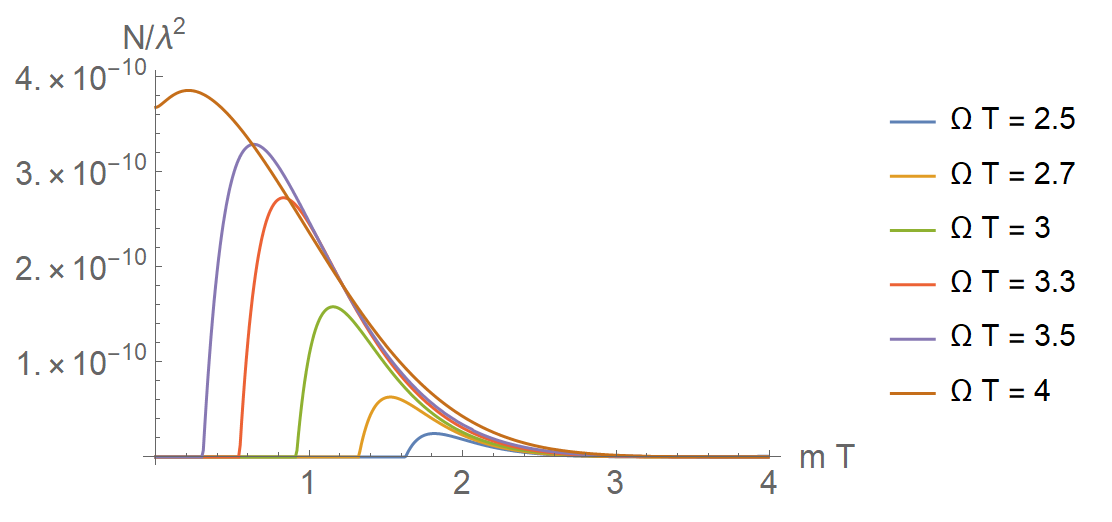}
        \caption{Negativity of the two detector system as a function of the field mass for different values of $\Omega$. We fixed the detector separation as $L = 5 T$ and the detector size as $\sigma = 0.1 T$.}\label{fig:scalar2}
    \end{figure}

    \begin{figure}[h]
        \includegraphics[scale=0.36]{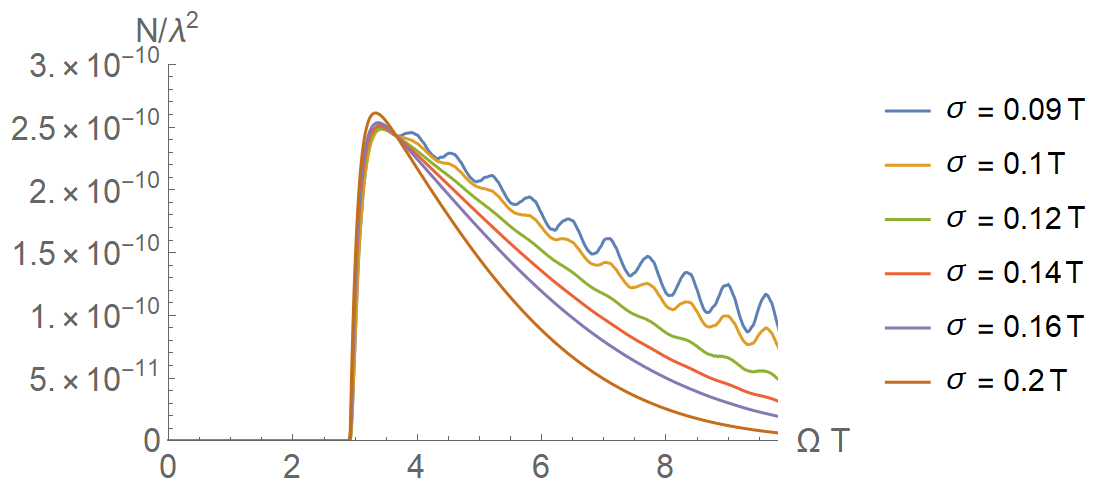}
        \caption{Negativity of the two detector system as a function of the detector gap for different values of $\sigma$. We fixed the detector separation as $L = 5 T$ and the field's mass as \mbox{$m T  = 1$}.}\label{fig:scalarOsc}
    \end{figure}

    \begin{figure}[h]
        \includegraphics[scale=0.4]{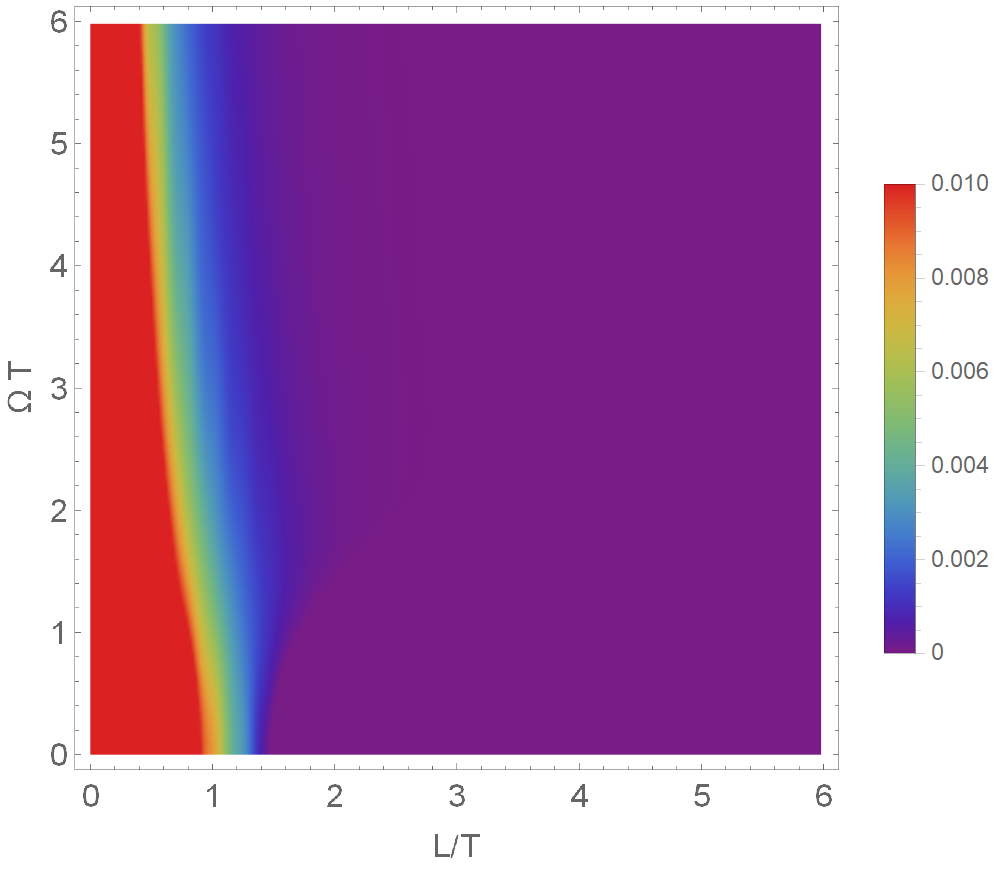}
        \caption{Negativity of the two detector system as a function of the detectors gap and separation. We fixed $m=0$ and considered pointlike detectors with $\sigma = 0$. }\label{fig:scalar2D}
    \end{figure}
    
    \begin{figure}[h]
        \includegraphics[scale=0.4]{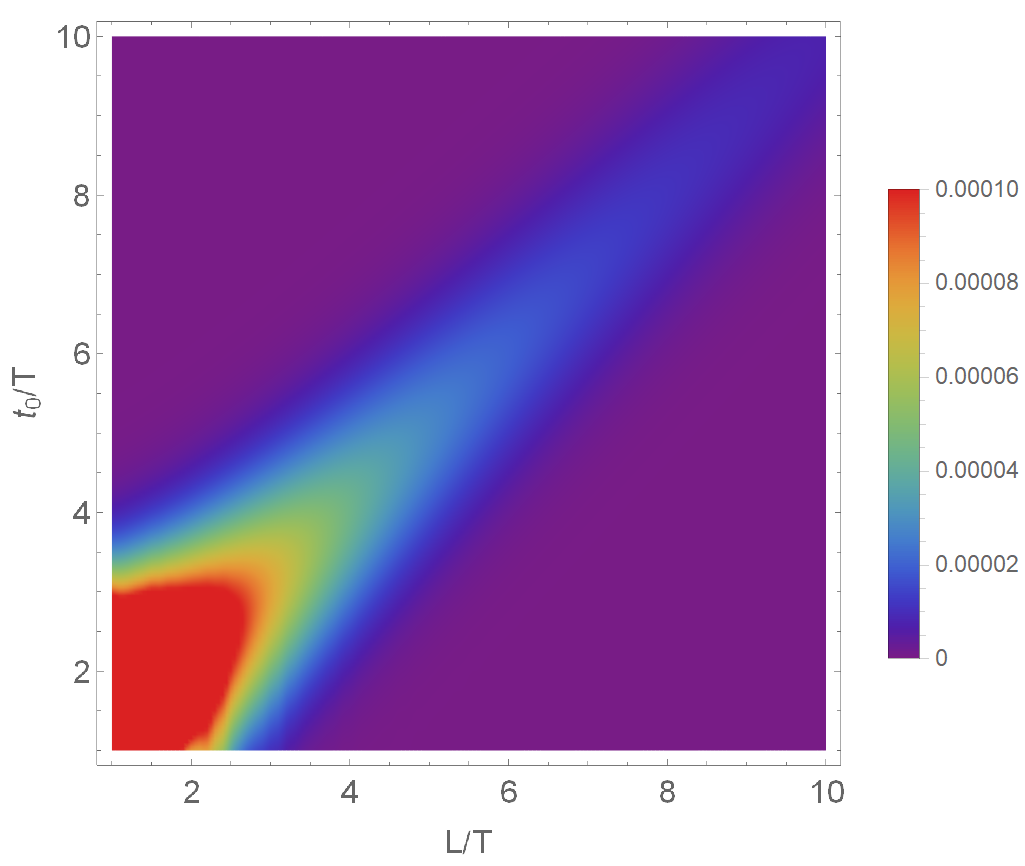}
        \caption{Negativity of the two detector system as a function of the detectors separation in space ($L$) and in time ($t_0$). We fixed $\Omega = 4/T$ for the detectors gap, $m=0$ for the field mass, and considered pointlike detectors with $\sigma = 0$. }\label{fig:scalar2DSpacetime}
    \end{figure}
    
    {In Figs.~\ref{fig:scalar} to~\ref{fig:scalar2DSpacetime} we plot the negativity of the two detectors state as a function of different parameters of the setup. \tb{Overall, we obtain that $\mathcal{N}^{(2)} /\lambda^2 \approx 10^{-10}$ for spacelike separated detectors, agreeing with the orders of magnitude for similar setups for scalar fields~\cite{Pozas-Kerstjens:2015,ericksonNew}}. In Fig. \ref{fig:scalar} we see the negativity as a function of the detectors' separation for different values of the field mass. Notice that as $L$ approaches $\sigma = 0.1 T$ the \tb{overlap between the detectors' spatial smearing becomes significant\footnote{\tb{Notice that rigorously speaking Gaussian smearings always overlap. However, for a separation beyond five standard deviations from the lightcone of the centre of the interaction of the detectors the effect on entanglement harvesting of the signalling between the detector tails is negligible beyond the computer precision used to perform the calculations so that the detectors can be considered effectively spacelike separated~\cite{Pozas-Kerstjens:2015,Pozas2016}}}}. We observe the following behaviour: the negativity decreases with the detector's separation, and for \tb{massive fields, the higher its mass}, the less entanglement can be extracted. \tb{The case of a massless field presents a different behaviour, where we see that the amount of entanglement that can be extracted from it is smaller than what can be extracted from a field with higher mass. It is not surprising that the massless case behaves different from the massive cases, as it is the only case where strong Hyugen's principle holds (see, e.g., \cite{Hyugens,Hyugens2,martin-martinez2015}).} \tbb{We remark that for $L\lesssim \sigma$ (that is, $L\lesssim 0.1T$) the plot represents an unphysical situation, where the detectors overlap, and that within the regime of this plot ($L<5T$), the negativity acquired by the detectors is mostly due to communication. 
    }
    
    In Figs. \ref{fig:scalar1} and \ref{fig:scalar2} we study the dependence of the entanglement harvested with the field mass $m$ and the detector gap $\Omega$ when the detectors separation is given by $L = 5T$, which guarantees that the spacetime regions of the interactions of the detectors are effectively spacelike separated (10$\sigma$ to 20$\sigma$ separation). It is possible to conclude the following: 1) As a function of the detector gap, there is a peak in negativity that is controlled by the value of the field mass: \tb{with the exception of the massless case, }the smaller the mass of the field, the higher the negativity peak.  2) Fields with higher mass allow entanglement to be harvested with smaller detector gaps. The overall behaviour of the negativity is similar to what is expected by the decay of \tb{the correlations of a quantum field with mass~\cite{birrell_davies}}. Notice that it is only possible to harvest entanglement if the detector gap is large enough.
    
    In Fig. \ref{fig:scalarOsc} we analyze the behaviour of the negativity as a function of the detectors gap  and size. For sufficiently large $\sigma$, the negativity \tb{peaks and after that becomes} monotonically decreasing with $\Omega$. On the other hand, as the detector size decreases, the negativity starts oscillating as a function of $\Omega$.
    
    Figure \ref{fig:scalar2D} shows the negativity as a function of both $L$ and $\Omega$ \tb{for pointlike detectors ($\sigma=0$)}. The region where it is possible to harvest entanglement, as well as the overall behaviour of $\mathcal{N}^{(2)}$ {presents differences when compared to} what is seen in the literature for the ground-ground protocol for entanglement harvesting~\cite{Pozas-Kerstjens:2015,Pozas2016,Petar,Cong2019,foo}. {\tb{It is well known in the literature that} in the ground-ground case, negativity vanishes for sufficiently small values of $\Omega T$, while \tb{we see here that} in the \tb{excited-ground} case, provided that the detectors are close enough (and definitely not spacelike separated~\cite{JormaPozas}), they are able to get entangled.} 
    Finally, in Fig. \ref{fig:scalar2DSpacetime} we consider \tb{pointlike detectors whose interactions are} separated by a time interval $t_0$ and by a distance $L$. Same as in the case of the ground-ground protocol, the most entanglement appears when the detectors' are within each other's lightcones, although in light contact most of the entanglement is not due to harvesting and rather due to direct communication~\cite{ericksonNew}. 

    }
    
    
    
    
    
    

\color{black}



    \section{{The Complex Scalar Particle Detector}}\label{sec:anti}
    
    As shown in~\cite{antiparticles}, it is possible to introduce a  modification to the original UDW detector model to study some properties of the anti-particle sector of non-Hermitian fields without giving up on the linearity of the coupling.
    Consider an $(n+1)$-dimensional spacetime \tb{$M$} with a complex scalar quantum field. The field $\hat{\psi}(\mf x)$ and its Hermitian conjugate, $\hat{\psi}^\dagger(\mf x)$ can be expanded in terms of a normalized basis of solutions of the Klein-Gordon equation, $\{u_{\bm p}(\mf x),u_{\bm p}^*(\mf x)\}$, according to
    \begin{align}
        \hat{\psi}(\mf x) &=\int \dd^n \bm p \left(u_{\bm p}(\mf x) \hat{a}_{\bm p}+u_{\bm p}^*(\mf x) \hat{b}^\dagger_{\bm p}\right),\label{eq:fieldComplex}\\
        \hat{\psi}^\dagger(\mf x) &= \int \dd^n \bm p \left(u_{\bm p}^*(\mf x) \hat{a}^\dagger_{\bm p}+u_{\bm p}(\mf x) \hat{b}_{\bm p}\right),
    \end{align}
    where $\hat{a}^\dagger_{\bm p}$ and $\hat{a}_{\bm p}$ are the creation and annihilation\vspace{-1mm}  
    operators  associated with particles and $\hat{b}^\dagger_{\bm p}$ and $\hat{b}_{\bm p}$ are the creation and annihilation operators associated with the anti-particle sector. 
    \tb{For a bosonic scalar field, the creation and annihilation operators satisfy the canonical commutation relations}
    \begin{align}
        \comm{\hat{a}^{\vphantom{\dagger}}_{\bm p}}{\hat{a}^\dagger_{\bm p'} } &= \delta^{(3)}(\bm p - \bm p'),&&& \comm{\hat{b}^{\vphantom{\dagger}}_{\bm p}}{\hat{b}^\dagger_{\bm p'} } &= \delta^{(3)}(\bm p - \bm p'),
    \end{align}
    \tb{with all other commutators vanishing.}

    Among all the possible couplings of particle detectors to complex scalar fields, to the author's knowledge, there are two proposals that have been studied in the literature:  quadratically~\cite{Sachs1,sachs2018entanglement,eduAchimBosonFermion} and  linearly coupled detectors~\cite{antiparticles}. In both cases, the detector's quantum system is defined along a timelike trajectory $\mf z(\tau)$, where $\tau$ is its proper time parameter. 
    
    The quadratic model considers a coupling with a Hermitian field observable that is quadratic on the field. The interaction Hamiltonian weight is prescribed as     \begin{equation}\label{hIQuad}
        \hat{h}_{I}(\mf x) =  \lambda\Lambda(\mf x)\hat{\mu}(\tau):\!\hat{\psi}^\dagger(\mf x)\hat{\psi}(\mf x)\!:,
    \end{equation} 
 where $\lambda$ is the coupling constant, $\Lambda(\mf x)$ is the \emph{real} spacetime smearing function and \mbox{$\hat{\mu}(\tau) = e^{\text{i}\Omega\tau}\hat{\sigma}^+ + e^{-\text{i}\Omega\tau}\hat{\sigma}^-$} is the detector's monopole moment. The normal ordering operation {in Eq. \eqref{hIQuad}} has been shown to be necessary to regularize spurious divergences~\cite{eduAchimBosonFermion}.  The response of this detector model to different field states and detector motions has been thoroughly studied in the literature~\cite{takagi}. In fact, in~\cite{Sachs1,sachs2018entanglement}, it has been shown that it is possible to use a pair of quadratically coupled particle detectors to harvest entanglement from the quantum vacuum.
    
    However, if one is interested in the study of the properties of the field associated to its anti-particle content, this model does not suffice. In fact, a disadvantage of the interaction given by Eq. \eqref{hIQuad} is that the detector couples equally to the particle and anti-particle sector of the field. This is a direct consequence of the fact that the operator $:\!\hat{\psi}^\dagger(\mf x)\hat{\psi}(\mf x)\!:$ is self-adjoint. This effectively makes the quadratically coupled detector unable to distinguish particles from anti-particles. 

   In turn, in the linear model, the detector has U(1) charge and therefore it can couple to the field and its complex conjugate in a U(1) invariant way~\cite{antiparticles}. 
   Moreover, the detector couples differently to $\hat{\psi}(\mf x)$ and $\hat{\psi}^\dagger(\mf x)$, so that the detector is able to distinguish particle and anti-particle content~\cite{antiparticles}. The Hamiltonian weight for this particle detector model is given by
    \begin{equation}\label{hIcomplex}
        \hat{h}_{I}(\mf x) =  \lambda\left(\Lambda^{\!(c)}(\mf x)e^{\mathrm{i}\Omega \tau}\hat{\sigma}^+\hat{\psi}^\dagger(\mf x) + \Lambda^{\!(c)}{}^*(\mf x)e^{-\mathrm{i}\Omega \tau}\hat{\sigma}^-\hat{\psi}(\mf x)\right),
    \end{equation}
    where $\Omega$ is the energy gap of the detector, $\hat{\sigma}^\pm$ are SU(2) ladder operators and $\Lambda^{\!(c)}(\mf x)$ is now a \emph{complex} spacetime smearing function. Once again, we denote the ground and excited states of the detector as $\ket{g}$ and $\ket{e}$, so that $\hat{\sigma}^+ = \ket{e}\!\!\bra{g}$ and $\hat{\sigma}^- = \ket{g}\!\!\bra{e}$.
    
    Particle detector models linearly coupled to complex fields have been thought to be unphysical~\cite{takagi}, specially due to the apparent break of the $U(1)$ symmetry present in the free field theory. However, in recent studies~\cite{antiparticles,neutrinos}, it has been shown that these model can be used to approximate the coupling of nucleons with the neutrino field. The $U(1)$ symmetry issue is fixed through the introduction of a complex degree of freedom in the detector itself. In fact, the complex smearing function $\Lambda^{\!(c)}(\mf x)$ must transform according to a $U(1)$ transformation, so that the full theory remains invariant~\cite{antiparticles}. \tb{In particular, the fact that the model is invariant under the transformations $\Lambda^{\!(c)}(\mf x)\longmapsto e^{\ii \alpha}\Lambda^{\!(c)}(\mf x)$, $\hat{\psi}(\mf x)\longmapsto e^{\ii \alpha}\hat{\psi}(\mf x)$ implies that any prediction of the model will also be invariant under this transformation. In particular, the results of entanglement harvesting presented in the next sections will all present this $U(1)$ invariance.}
    
    \tbb{When the field is in a zero-mean Gaussian, the final state of the detector can be entirely determined by the field two-point function $\langle \hat{\phi}(\mf x) \hat{\phi}(\mf x')\rangle$. In fact, all measurable quantities involving the field should be expressible in terms of $\langle \hat{\phi}(\mf x) \hat{\phi}(\mf x')\rangle $ since for zero-mean Gaussian states all $n$-point functions are a function of the two-point function.} In particular, for the linear complex detector model interacting with the vacuum, the final state of the detector will depend on the following Wightman functions
    \begin{align}\label{antiCorrVac}
        \bra{0}\!\hat{\psi}(\mf x')\hat{\psi}^\dagger(\mf x)\!\ket{0}\! &= \!\bra{0}\!\hat{\psi}^\dagger(\mf x')\hat{\psi}(\mf x)\!\ket{0} \!=\! \int \dd^n\bm k u_{\bm k}(\mf x') u_{\bm k}^*(\mf x),
    \end{align}
    because $        \bra{0}\!\hat{\psi}(\mf x')\hat{\psi}(\mf x)\!\ket{0}\! = \!\bra{0}\!\hat{\psi}^\dagger(\mf x')\hat{\psi}^\dagger(\mf x)\!\ket{0} =0.$ Notice that the non-zero two-point functions above are the same as we had in the real scalar case in Eq. \eqref{wPhi}. In particular, this implies that the excitation probability of a single detector interacting linearly with a complex field in the vacuum state will be the same as if it were interacting linearly with a real scalar field. In fact, this was explicitly shown in \cite{antiparticles}, where these models were compared and the differences only arise when the field is in states that have non-zero particle {or} anti-particle content. 
    
    On the other hand, the entanglement harvesting capability of detectors that are linearly coupled to non-Hermitian fields has not yet been studied in the literature to the author's knowledge. The model studied in this section is of particular importance, especially due to its similarities with the well-known UDW model. This allows for a comparison of the entanglement that can be harvested from the vacuum for non-Hermitian and real fields that is not polluted by the non-linear nature of the coupling.

\section{Entanglement Harvesting from complex fields with linearly \\ coupled detectors}\label{sec:antiHarvest}

    In this section we explore an entanglement harvesting setup when two detectors couple to a complex scalar field linearly, according to the interaction Hamiltonian weight in Eq. \eqref{hIcomplex}. That is, we consider two anti-particle UDW detectors labelled by $j=1,2$ undergoing time-like trajectories $\mf z_j(\tau_j)$, where $\tau_j$ denotes their respective proper times. The interaction Hamiltonian weights between the detectors and the field can be written as
    \begin{equation}\label{hIcomplexj}
        \hat{h}_{I,j}(\mf x) \!= \! \lambda\!\left(\!\Lambda^{\!(c)}_j\!(\mf x)e^{\mathrm{i}\Omega_j \tau_j}\hat{\sigma}_j^+\hat{\psi}^\dagger(\mf x) \!+\! \Lambda^{\!(c)*}_j\!(\mf x)e^{-\mathrm{i}\Omega_j \tau_j}\hat{\sigma}_j^-\hat{\psi}(\mf x)\!\right)\!,
    \end{equation}
    where $\Omega_j$ is the $j$-th detector's energy gap, $\hat{\sigma}^\pm_j$ are the respective two-level raising and lowering operators and $\Lambda^{\!(c)}_{j}\!(\mf x)$ are the spacetime smearing functions associated with each detector. The full interaction Hamiltonian weight that governs the interaction will then be given by the sum of the individual interactions for $j=1,2$:
    \begin{equation}\label{hIcomplexsum}
        \hat{h}_{I}(\mf x) =  \hat{h}_{I,1}(\mf x)+\hat{h}_{I,2}(\mf x).
    \end{equation}
    
    As we mentioned in Section \ref{sec:anti}, where the anti-particle detector model was reviewed, these detectors couple differently to the particle and anti-particle sector of the quantum field theory: the ground state of the detector couples to the anti-particle sector, while the excited state couples to the particle sector. This has meaningful consequences for the entanglement harvesting setup, and {motivates the division of} our analysis between the case where the detectors are both in the ground state and the case where the detectors start in different states. In the next two subsections we will discuss these two cases in detail.

\subsection{Ground - Ground Protocol}\label{sub:antiHarvestgg}

    We first assume that both detectors  start in their respective ground states. Assuming the field to start in a state $\hat{\rho}_{\phi}$, we can then write the full detectors-field initial state as
    \begin{equation}
        \hat{\rho}_0 = \ket{g_1}\!\!\bra{g_1}\otimes\ket{g_2}\!\!\bra{g_2}\otimes \hat{\rho}_{\phi} = \hat{\rho}_{\text{d}, 0}\otimes \hat{\rho}_{\phi}.
    \end{equation}
    Using the time evolution operator of Eq. \eqref{UI}, with the interaction Hamiltonian weight in Eq. \eqref{hIcomplexsum}, it is possible to obtain the final state of the detectors quantum system. We perform the explicit computations in Appendix \ref{app:gg}, and obtain the following density matrix     
    \begin{equation}
        \hat{\rho}_{\text{d}} = \begin{pmatrix}
            1 - \L_{11}-\L_{22} & \mathrm{i}\mathcal{E}_2\color{black} & \mathrm{i}\mathcal{E}_1\color{black} & \mathcal{M}^*\\
             -\mathrm{i}\mathcal{E}_2^* & \L_{22} & \L_{21} & 0\\
            -\mathrm{i}\mathcal{E}_1^* & \L_{12} & \L_{11} & 0\\
            \mathcal{M} & 0 & 0 & 0 
        \end{pmatrix} + \mathcal{O}(\lambda^3),
    \end{equation}
    where the matrix elements are given by    
    \begin{align}
        \L_{ij} &=\lambda^2\int \dd V \dd V' e^{\mathrm{i}\Omega_i \tau_i-\mathrm{i}\Omega_j \tau_j'}\langle\hat{\psi}_i(\mf x')\hat{\psi}_j^\dagger(\mf x) \rangle_{{\rho}_{\phi}},\nonumber\\ 
        \mathcal{M} &= - \lambda^2 \int \dd V\dd V'\theta(t-t')\nonumber\\&\:\:\:\:\:\:\:\:\:\:\:\:\:\:\:\:\Big(e^{\mathrm{i}\Omega_1 \tau_1 + \mathrm{i}\Omega_2 \tau_2'}\langle\hat{\psi}_1^\dagger(\mf x)\hat{\psi}_2^\dagger(\mf x')\rangle_{\rho_{\phi}}\nonumber\\
        &\:\:\:\:\:\:\:\:\:\:\:\:\:\:\:\:\:\:\:\:\:\:\:\:\:\:\:\:\:\:\:\:+e^{\mathrm{i}\Omega_2 \tau_2 +\mathrm{i}\Omega_1 \tau_1'}\langle\hat{\psi}_2^\dagger(\mf x)\hat{\psi}_1^\dagger(\mf x')\rangle_{\rho_{\phi}}\Big),\nonumber\\
        \mathcal{E}_i &    = \lambda \int \dd V  e^{\tbb{-}\mathrm{i} \Omega_i \tau_i} \langle\hat{\psi}_i(\mf x)\rangle_{\rho_{\phi}},\label{antiexpgg}
    \end{align}
    and we have defined
    \begin{equation}\label{antiPsii}
        \hat{\psi}_i(\mf x) \coloneqq \Lambda^{(c)*}_i\!(\mf x) \hat{\psi}(\mf x), \quad\quad \hat{\psi}_i^\dagger(\mf x) \coloneqq \Lambda^{(c)}_i\!(\mf x) \hat{\psi}^\dagger(\mf x).
    \end{equation}
    Notice the similarity with the case of the standard UDW model that couples to a real scalar field (Eq.~\eqref{expReal}). Namely, with the replacement $\hat{\psi}(\mf x) \mapsto \hat{\phi}(\mf x)$ and assuming $\Lambda^{\!(c)}_j\!(\mf x)$ to be real, we recover the exact expressions from Subsection \ref{sub:UDWharvest}. In particular, the negativity of the state $\hat{\rho}_{\text{d}}$ is given by the same expression that we had for the scalar case in Eq.~\eqref{UDWN}. \tb{In the case when $\mathcal{E}_1 = \mathcal{E}_2 = 0$, the negativity will only be} non-zero if the $\mathcal{M}$ component is positive and larger than the product $\mathcal{L}_{11}\mathcal{L}_{22}$.
    
    Despite this similarity, there are stark differences in the ability of the detector to harvest entanglement in the real and the complex scalar field. 
    Indeed, for a complex scalar field we have
    \begin{equation}\label{eq:noSelf}
        \langle \hat{\psi}(\mf x) \hat{\psi}(\mf x') \rangle_0  = \langle \hat{\psi}^\dagger(\mf x) \hat{\psi}^\dagger(\mf x') \rangle_0 = 0.
    \end{equation}
    In particular, this implies that, to leading order, a pair of detectors that start in the ground state cannot harvest entanglement from the vacuum of a complex scalar field when linearly coupled according to the interaction in Eq.~\eqref{hIcomplexj}. This can be intuitively understood {by noticing} that for charge conjugation invariant states, the field correlations appear only between the particle and anti-particle sectors. When both detectors start in the ground state, they couple only to the anti-particle sector, which does not contain self-correlations as per Eq.~\eqref{eq:noSelf}. In fact, as we will see in the next Subsection, when one detector couples to the particle content and the other one couples to the anti-particle content of the field, it is possible to extract entanglement from the vacuum state. Notice that this ground-ground inability to harvest entanglement from the charge conjugation-invariant states at leading order is quite general and holds in curved spacetimes and independently of the Hilbert space representation of the field.
    
    
    Finally, notice that although we have restricted our analysis to the case where both detectors start in the ground state, the same would apply to the case where both detectors start in the excited state. In fact, as discussed in~\cite{antiparticles}, \tb{the excited case can be obtained from the ground case by changing the sign of the gaps (\mbox{$\Omega_i \longrightarrow -\Omega_i$}) and conjugating the smeared fields (\mbox{$\hat{\psi}_i(\mf x)\mapsto \hat{\psi}_i^\dagger(\mf x)$}). For more details, we refer the reader to the discussion in the end of Appendix \ref{app:ge}}. As a conclusion, if both detectors start in the ground state or the excited state, they are unable to harvest entanglement from any charge-conjugation invariant state \tb{(e.g. the vacuum) due to the vanishing of the two point functions $\langle \hat{\psi}(\mf x) \hat{\psi}(\mf x') \rangle$ and $\langle \hat{\psi}^\dagger(\mf x) \hat{\psi}^\dagger(\mf x') \rangle$}. 
    
    
\subsection{\tbb{Excited - Ground} Protocol}\label{sub:antiharvestge}

    We now consider the case where one detector starts in the ground state, coupled to the anti-particle sector of the field, while the other detector starts in the excited state, coupled to the particle sector. In this setup we can write the initial state of the detectors-field system, $\hat{\rho}_0$, in an analogous form to~\eqref{eq:geinitialscalar} as
    \begin{equation}
        \hat{\rho}_0 = \ket{e_1}\!\!\bra{e_1}\otimes\ket{g_2}\!\!\bra{g_2}\otimes \hat{\rho}_{\phi} = \hat{\rho}_{\text{d}, 0}\otimes \hat{\rho}_{\phi}.
    \end{equation}
    
    Performing the computations in the same fashion as we did in Subsection \ref{sub:UDWharvestge}, we obtain the final density operator associated with the two detectors, after tracing out the field degrees of freedom. The computations can be found in Appendix \ref{app:ge}, where we obtain the density matrix for the detectors system:
    \begin{equation}\label{rhoDge}
        \hat{\rho}_{\text{d}} =\begin{pmatrix}
            \tb{\mathcal{L}_{11}'} & 0 & -\mathrm{i}\mathcal{E}_1 & \tb{\mathcal{L}_{12}'}\\
            0 & 0 & \mathcal{M}' & 0 \\
            \mathrm{i}\mathcal{E}_1^* & \mathcal{M}'^* & 1 - \tb{\mathcal{L}_{11}'} - \mathcal{L}_{22} & \mathrm{i}\mathcal{E}_2\\
            \tb{\mathcal{L}_{12}'^*} & 0 & -\mathrm{i}\mathcal{E}_2^* & \mathcal{L}_{22}        
        \end{pmatrix}+ \mathcal{O}(\lambda^3),
    \end{equation}
    where $\mathcal{L}_{ii}$ and $\mathcal{E}_i$ are given by Eq. \eqref{antiexpgg} and the remaining components are explicitly given by
    \begin{align}
        \tb{\mathcal{L}_{12}'} &= \lambda^2 \int \dd V \dd V' e^{-\mathrm{i} \Omega_1 \tau_1 -\mathrm{i}\Omega_2 \tau_2'} \langle\hat{\psi}_1(\mf x) \hat{\psi}_2(\mf x')\rangle_{\hat{\rho}_\phi}\label{Nijc},\\
        \mathcal{M}' &= \lambda^2  \int \dd V \dd V' \theta(t-t')\nonumber\\*
        &\:\:\:\:\:\:\:\:\:\:\:\:\:\:\:\times\Big(e^{\mathrm{i}(\Omega_2 \tau_2'-\Omega_1 \tau_1)}\!\langle \hat{\psi}_1(\mf x)\hat{\psi}_2^\dagger(\mf x') \rangle_{\rho_{\phi}}\nonumber\\*
        &\:\:\:\:\:\:\:\:\:\:\:\:\:\:\:\:\:\:\:\:\:\: +e^{\mathrm{i}(\Omega_2 \tau_2 -\Omega_1 \tau_1')}\!\langle\hat{\psi}_2^\dagger(\mf x)\hat{\psi}_1(\mf x')\rangle_{\rho_{\phi}} \!\Big),\nonumber\\
         {\tb{\mathcal{L}_{11}'}} &{= \lambda^2\int \dd V \dd V' e^{-\ii \Omega_1  ({\tau_1-\tau_1}')}\langle \hat{\psi}_{\tbb{1}}(\mf x)\hat{\psi}^\dagger_{\tbb{1}}(\mf x')\rangle_{\rho_{\phi}},} \nonumber
    \end{align}
    where we are using the notation from Eq. \eqref{antiPsii} for $\hat{\psi}_j(\mf x)$. Same as we had in the real scalar case, the $\mathcal{M}'$ term is slow rotating, which simplifies the numerical evaluation of the negativity.
    
    Specifically, the negativity for the case where \mbox{$\mathcal{E}_1 = \mathcal{E}_2 = 0$} is given by Eq. \eqref{negrge}, where now $\mathcal{M}'$ and $\tb{\mathcal{L}_{12}'}$ are given by Eq. \eqref{Nijc}. Namely, 
    \begin{equation}
        \mathcal{N}^{(2)} = \max\left(0,\sqrt{{|\mathcal{M}'|}^2+\frac{(\tb{\L_{11}'} -\L_{22})^2}{4} }-\frac{\tb{\mathcal{L}_{11}'} + \mathcal{L}_{22}}{2}\right),
    \end{equation}
    where it should be noted that the $\mathcal{M}'$ term now involves different field correlators, that couple the particle and anti-particle sector of the field. In particular, using the fact that the complex and \tb{real} scalar correlators in the vacuum state are the same (Eq. \eqref{antiCorrVac}), it is easy to see that this detector model can harvest \tb{at least} as much entanglement as the detector that couples to the real scalar field, since the $\tb{\mathcal{L}_{ij}'}$ and $\mathcal{M}'$ terms are the same, and therefore, so is the negativity. 
    
    At last, notice that the explicit examples of Subsection \ref{sub:UDWharvestge} yield the same results as they would for the linearly coupled detectors in this case by choosing \tb{the complex spacetime smearing functions to be real}. This is due to the fact that the $\tb{\mathcal{L}_{12}'}$ terms do not contribute to the negativity, and the fact that
    \begin{equation}\label{propagators}
        \bra{0}\!\hat{\psi}(\mf x')\hat{\psi}^\dagger(\mf x)\!\ket{0}\! = \!\bra{0}\!\hat{\psi}^\dagger(\mf x')\hat{\psi}(\mf x)\!\ket{0} \!=\!\bra{0}\!\hat{\phi}(\mf x')\hat{\phi}(\mf x)\!\ket{0}.
    \end{equation}
    That is, all conclusions drawn from the explicit example in Subsection \ref{sub:UDWharvestge} are also valid for the linear complex particle detector models when choosing $\Lambda^{\!(c)}\!{(\mf x)}  = \Lambda{(\mf x)}$.

    \color{black}
    
    \subsection{Enhancing entanglement harvesting with a complex smearing function}
    
    Here we will study the impact on entanglement harvesting considering complex smearing functions. The only effect of adding a complex phase to any smearing function $F(\bm x)\mapsto F(\bm x)e^{\ii \bm a \cdot \bm x}$ is shifting the Fourier transform of the smearing appearing in all our expressions, so studying the effects of such phase can be quickly done with the expressions already computed.
    
    To illustrate, we will consider an explicit example of entanglement harvesting from a complex scalar field using a complex smearing function. As we will see, using a complex smearing function allows the detectors to harvest a significantly larger amount of entanglement. 
    
    We consider two identical inertial detectors in Minkowski spacetime, according to the interactions of Eq. \eqref{hIcomplexj}. We consider the detectors' gap to be $\Omega$, and their space separation to be $\bm L$ with $L = |\bm L|$, so that their respective spacetime smearing functions can be written as 
    \begin{align}
        \Lambda_1(\mf x) &= \chi(t) F_{\bm a}(\bm x),\\
        \Lambda_2(\mf x) &= \chi(t) F_{\bm a}(\bm x-\bm L),
    \end{align}
    where
    \begin{equation}
        F_{\bm a}(\bm x) = F(\bm x)e^{\ii \bm a \cdot\bm x}
    \end{equation}
    and $F(\bm x)$ is a \emph{real} smearing function. Then the phase $e^{\ii \bm a \cdot \bm x}$ determines the complex degree of freedom of the smearing function. Physically, the parameter $\bm a$ can be interpreted analogously to an electric dipole in the case of a vector interaction, where the direction of the dipole vector has meaningful consequences to the interaction of detectors with the field~\cite{Pozas2016}.
    
    The consequence of adding a complex phase $e^{\ii \bm a \cdot \bm x}$ to the smearing function can also be interpreted as a shift in momentum space. In fact, we have
    \begin{equation}
        \tilde{F}_{\bm a}(\bm p) = \tilde{F}(\bm p + \bm a).
    \end{equation}
    Then, with the specific choices we have made, the $\mathcal{M}$ and $\mathcal{L}$ terms relevant for entanglement harvesting can be written as
    \begin{align}
        &\mathcal{L}{(\Omega)} = \frac{\lambda^2}{(2\pi)^n}\!\int \frac{\dd^n \bm p}{2 \omega_{\bm p}}|\tilde{F}(\bm p+\bm a)|^2 |\tilde{\chi}( \omega_{\bm p}+\Omega)|^2,\label{Lcomplexss}\\
        &\mathcal{M}' =\frac{\lambda^2}{(2\pi)^n} \!\int \!\frac{\dd^n \bm p}{2 \omega_{\bm p}} |\tilde{F}(\bm p+\bm a)|^2 e^{\mathrm{i} \bm p \cdot \bm L}\label{Mcomplexss}\\&\:\:\:\:\:\:\:\:\:\:\:\:\:\:\:\:\:\:\:\:\:\:\:\:\:\:\:\:\:\:\:\:\:\:\times\left(Q(\Omega-\omega_{\bm p})+Q(-\omega_{\bm p}-\Omega)\right),\nonumber
    \end{align}
    where $Q(\omega)$ is defined in terms of $\chi(t)$ by Eq. \eqref{Comega}. By looking at the expressions above, we conclude that it is possible to shift the smearing function $|\tilde{F}(\bm p)|$. This can have noticeable consequences for entanglement harvesting. If the maximum of the Fourier transform of the smearing function happens for a momentum $\bm p$ so that $\omega_{\bm p}\approx \Omega$ we can maximize the amount of entanglement harvested. It is indeed common to consider cases where $\tilde{F}(\bm p)$, $\tilde{\chi}(\omega)$ and $\tilde{Q}(\omega)$ have well defined peaks, so that by regulating the parameter $\bm a$, we can make those functions peak on resonance, maximizing the integrals of Eqs. \eqref{Lcomplexss} or $\eqref{Mcomplexss}$.

    In order to further explore this phenomenon let us focus on a concrete example. In particular let us analyze an example analogous to that of Subsection \ref{sub:egex}: we pick Gaussian switching functions as in Eq. \eqref{chi}, and consider \emph{complex} smearing functions given by
    \begin{equation}
        F_{\bm a}(\bm x) = \frac{1}{(2\pi \sigma^2)^\frac{n}{2}} e^{-\frac{\bm x^2}{2 \sigma^2}} e^{\ii \bm a \cdot \bm x},
    \end{equation}
    In this case the Fourier transform of the complex smearing function reads
    \begin{equation}
        \tilde{F}_{\bm a}(\bm p) = e^{-\frac{\sigma^2(\bm p + \bm a)^2}{2}}.
    \end{equation}
    With the expression above, and the definition of $Q(\omega)$ from Eq. \eqref{expC}, we obtain the following expressions for $\mathcal{L}(\Omega)$ and $\mathcal{M}'$:
    \begin{align}
        \mathcal{L}{(\Omega)} = \frac{\lambda^2}{2\pi^2} \int_0^\infty \frac{\dd |\bm  p|}{2 \omega_{\bm p}} &|\bm p|^2 \:e^{-\sigma^2(|\bm p|^2 + |\bm a|^2)} e^{-(\omega_{\bm p} + \Omega)^2T^2} \nonumber\\
        &\:\:\:\:\:\:\:\:\:\:\:\:\:\:\:\:\:\:\:\:\:\:\:\:\times \frac{\sinh\left(2 |\bm p|\,|\bm a|\sigma^2\right)}{2|\bm p|\,|\bm a|\sigma^2}\nonumber,\\
        \mathcal{M}'
        =\frac{\lambda^2}{{4\pi^2}} \int_0^\infty \frac{\dd |\bm  p|}{2 \omega_{\bm p}} &|\bm p|^2 \:e^{-\sigma^2(|\bm p|^2 + |\bm a|^2)}\label{complexresult}\\ &\!\!\!\!\!\!\!\!\!\!\!\!\!\!\times\frac{\sinh\left( |\bm p| \sqrt{4|\bm a_\perp|^2 \sigma^4 - (\bm L-\ii\bm a_\parallel)^2}\right)}{|\bm p| \sqrt{4|\bm a_\perp|^2 \sigma^4 - (\bm L-\ii\bm a_\parallel)^2}}\nonumber\\&\times \Big(Q(\Omega-\omega_{\bm p})+Q(-\omega_{\bm p}-\Omega)\Big).\nonumber       
    \end{align}
    In the expressions above we decomposed $\bm a = \bm a_\perp + \bm a_\parallel$, where $\bm a_\parallel$ is the component of $\bm a$ in the direction of $\bm L$, and $\bm a_{\perp}$ is its component orthogonal to $\bm L$. For simplicity, we choose the $z$ axis parallel to $\bm L$ and the $x$ axis parallel to $\bm a_\perp$, so that $\bm a$ has no component in the $y$ direction, and can be written as $\bm a = a_x \bm e_x + a_z \bm e_z$.
    
\begin{figure}[h]
        \includegraphics[scale=0.4]{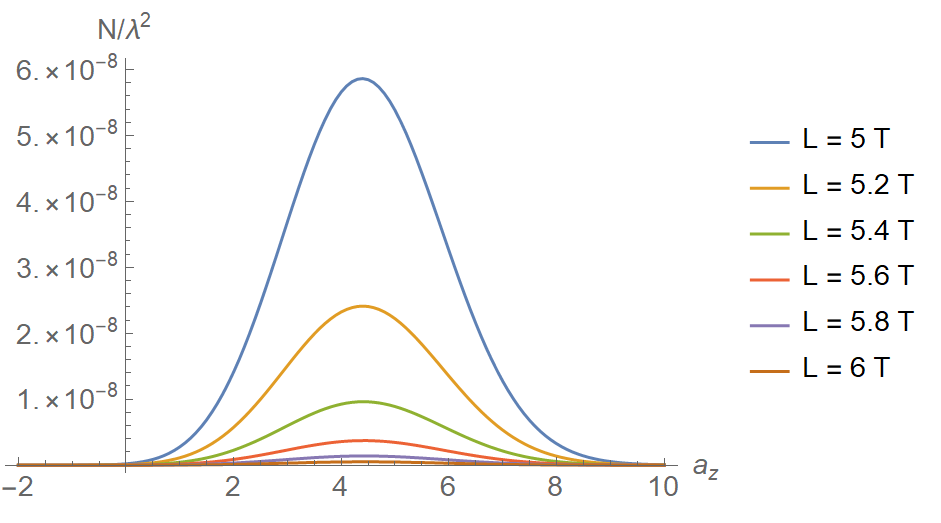}
        \caption{\tb{Negativity of the two detector system as a function of the parallel component of the vector parameter $\bm a$ for varying $L$ with $a_x = 0$. We fixed \tbb{the field's mass so that} $m T =0.5$, \tbb{the detectors' gap so that} $\Omega T = 4$ and considered \tbb{the detectors' size} $\sigma = 0.5 T$.}}\label{fig:azL}
    \end{figure}

\begin{figure}[h]
        \includegraphics[scale=0.4]{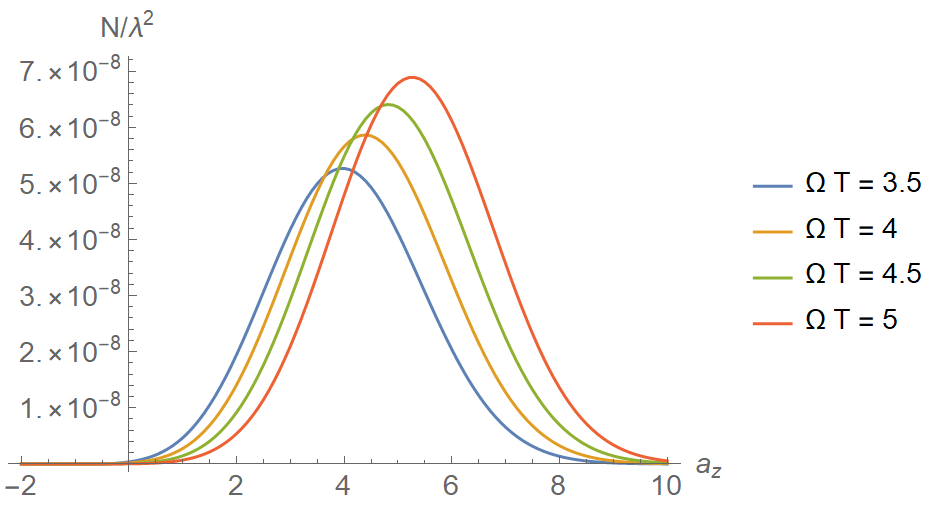}
        \caption{\tb{Negativity of the two detector system as a function of the parallel component of the vector parameter $\bm a$ for varying $\Omega$ with $a_x = 0$. We fixed \tbb{the field's mass so that} $m T =0.5$, \tbb{the detectors' separation so that} $L = 5T$ and considered \tbb{the detectors' size} $\sigma = 0.5 T$.}}\label{fig:azOmega}
    \end{figure}

\begin{figure}[h]
        \includegraphics[scale=0.41]{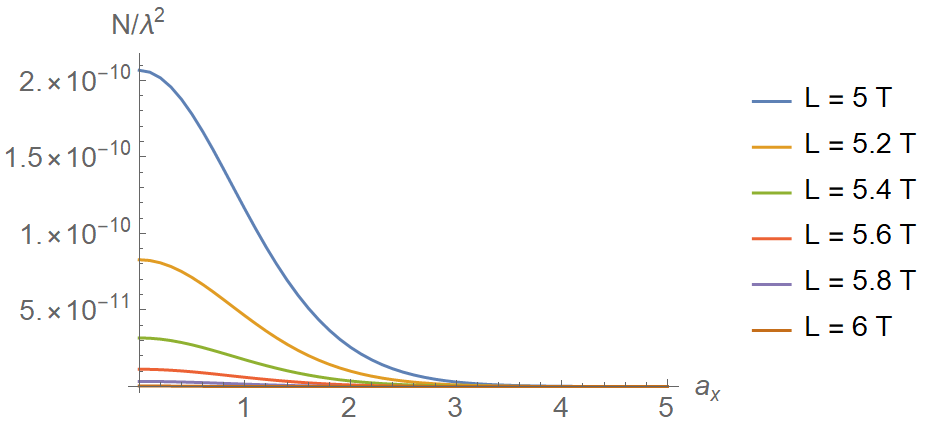}
        \caption{\tb{Negativity of the two detector system as a function of the orthogonal component of the vector parameter $\bm a$ for varying $L$ with $a_z = 0$. We fixed \tbb{the field's mass so that} $m T =0.5$, \tbb{the detectors' gap so that} $\Omega T = 4$ and considered \tbb{the detectors' size} $\sigma = 0.5 T$.}}\label{fig:axL}
    \end{figure}

\begin{figure}[h]
        \includegraphics[scale=0.4]{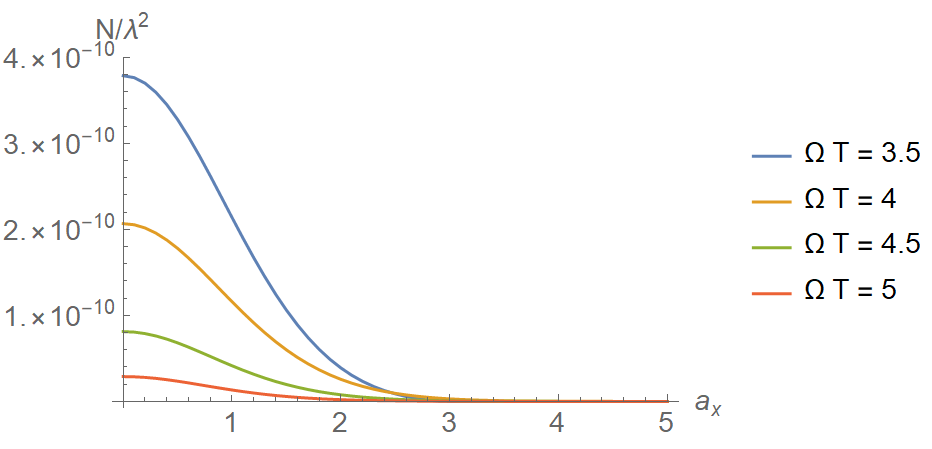}
        \caption{\tb{Negativity of the two detector system as a function of the orthogonal component of the vector parameter $\bm a$ for varying $\Omega$ with $a_z = 0$. We fixed \tbb{the field's mass so that} $m T =0.5$,  \tbb{the detectors' separation so that} $L = 5T$ and considered \tbb{the detectors' size} $\sigma = 0.5 T$.}}\label{fig:axOmega}
    \end{figure}

    We are interested in the behaviour of the negativity as a function of $\bm a$, given that it is the parameter that controls the complex degree of freedom of the smearing function $F(\bm x)$. In Figs. \ref{fig:azL}, \ref{fig:azOmega}, \ref{fig:axL} and \ref{fig:axOmega}, we plot the negativity as a function of the parameters $a_x$ and $a_z$ for different values of $\Omega$ and $L$, always ensuring that the detectors' interaction regions are approximately spacelike separated. Overall, we see that varying $\bm a$ can make the negativity both increase or decrease, as detailed below.
  
    In Figs. \ref{fig:azL} and \ref{fig:azOmega} we plot the negativity as a function of the component of $\bm a$ parallel to $\bm L$, $a_z$, for different values of $L$ and $\Omega$. Overall, we find that the negativity can either increase or decrease as the parameter $a_z$ varies. As expected, we find that the negativity decays as the separation between the detectors increases.  In Fig. \ref{fig:azOmega} we plot the negativity for different values of $\Omega$ keeping a separation of $L = 5 T$. We see that the values of $a_z$ that maximize the negativity are close to the gap $\Omega$, giving rise to the resonance behaviour predicted after Eq.~\eqref{Mcomplexss}.
    
    In Figs. \ref{fig:axL} and \ref{fig:axOmega} we plot the negativity as a function of the component of $\bm a$ orthogonal to $\bm L$, $a_x$, for varying $L$ and $\Omega$. Notice that the perpendicular component only depends on $|\bm a_\perp|$, so we only plot for positive values of $a_x$. Overall, we see that the negativity {decreases as the component $a_x$ increases. That is, we find that the orthogonal component of the vector $\bm a$ to the detectors separation does not help the harvesting of entanglement from a complex scalar field.} In Fig. \ref{fig:axL}, we see that as $L$ increases, the negativity monotonically decreases, as expected in entanglement harvesting. In Fig. \ref{fig:axOmega} we plot the negativity for varying values of $\Omega$, and we find that as $\Omega$ increases, the negativity decreases.
    
    We then see that adding a position dependent complex phase to the spacetime smearing function can increase the harvested entanglement by up to two orders of magnitude. \tbb{Moreover, although we have considered complex smearing functions for linearly coupled particle detector models, our results are also valid for the case of a linear coupling with a real scalar field that uses complex smearing functions. In fact, \tbbb{consider} the interaction Hamiltonian weight}
    \begin{equation}\label{MYREFEREE}
        \tbb{\hat{h}_I(\mf x) = \lambda \left(\Lambda^{\!(r)}(\mf x) e^{\ii \Omega\tau}\hat{\sigma}^{+}+\Lambda^{\!(r)*}(\mf x)e^{-\ii \Omega\tau} \hat{\sigma}^{-}\right)\hat{\phi}(\mf x)},
    \end{equation}
    \tbb{where $\Lambda^{(r)}(\mf x)$ is a \emph{complex} spacetime smearing function and $\hat{\phi}(\mf x)$ is a real scalar quantum field. Using Eq. \eqref{propagators}, we see that all results from this section also apply to the interaction~\eqref{MYREFEREE}. As far as the authors are aware, this model has never been considered for entanglement harvesting, but we showed in this section that such a model would have an advantage of two orders of magnitude over entanglement harvesting protocols that consider real detectors.}
    
    \color{black}

\section{The Fermionic Particle Detector}\label{sec:fermion}

    In this section we review the physical motivation behind the fermionic particle detector model introduced in~\cite{neutrinos}. We then consider the more general fermionic detector that was studied in~\cite{antiparticles}, and use it in an entanglement harvesting setup.
    
    \subsection{The Fermi Interaction}\label{sub:4fermi}
    
    In \cite{neutrinos}, it was shown that the interaction of nucleons with the neutrino fields through a four-fermion interaction can be modelled by a particle detector model. The four-fermion theory {effectively describes} the interaction of neutrons, protons, electrons and neutrinos by treating these as spinor fields $n,p,e$ and $\nu_e$, respectively. The interaction is prescribed in terms of the Lagrangian density    
    \begin{equation}\label{eq:4FLag}
        \mathcal{L}_{\text{4F}} = -\frac{ G_F}{\sqrt{2}}\Big(\left(\bar{\nu}_e\gamma^\mu P_Le\right)\left(\bar{n}\gamma_\mu P_Lp\right) + \text{H.c.}\Big),
    \end{equation}
    where $G_F \approx  1.16 \times 10^{-5}\, \text{GeV}^{-2}$ is the Fermi constant and $P_L$ denotes the projector in the left-handed component of the spinors, where the weak interaction takes place. This model is especially useful for the description of the $\beta^-$-decay, where a neutron decays into a proton, electron and anti-neutrino,
    \begin{equation}\label{betadecay}
        n \longrightarrow p+e^- + \bar{\nu}_e.
    \end{equation}

    The procedure to reduce the model from Eq. \eqref{eq:4FLag} to a localized particle detector comes in two steps. First we notice that the neutron and proton masses are much larger than both the electron and neutrino masses. Using the fact that the neutron and proton are always coupled to each other in the 4 fermion Lagrangian, we then replace their sector by a single two-level system, according to    
    \begin{equation}\label{eq:pn}
        \bar{n}\gamma_\mu P_Lp \longrightarrow j_{\mu}(\mf x)e^{\mathrm{i}\Delta M \tau}\ket{n}\!\!\bra{p},
    \end{equation}
    where we denote the neutron and proton states by $\ket{n}$ and $\ket{p}$, respectively. In Eq. \eqref{eq:pn}, $\Delta M$ is the mass difference of the neutron and proton and $j^\mu(\mf x)$ is a hadronic current, that gives the spatial profile of the neutron/proton system.
    
    The next step is to notice that the electron mass is of the order of $10^5$ times larger than the neutrino mass, so that the electrons involved in the process are considered to be localizable in terms of their interaction with the neutrino field (the interaction happens in a finite region of spacetime). We then make the following replacement in Eq. \eqref{eq:4FLag},
    \begin{equation}\label{eq:el}
        P_L\hat{e} \longrightarrow u(\mf x)e^{-\mathrm{i}\omega_e \tau} \hat{a}_e,
    \end{equation}
    where $\omega_e$ is the energy scale of the electrons involved in the specific process considered, $u(\mf x)$ is an effective mode associated with the spatial profile of the electron and $\hat{a}_e$ is the creation operator associated with the electron field that participates in the interaction.

   
    
    With the two replacements of Eqs. \eqref{eq:pn} and \eqref{eq:el} in Eq. \eqref{eq:4FLag}, we end up with the following effective Lagrangian, 
    \begin{align}
        \mathcal{L}_{\text{4F}} &= -\dfrac{G_F}{\sqrt{2}}\Big(e^{\mathrm{i}(\Delta M - \omega_e)\tau}\bar{\nu}_e\slashed{j}(\mf x)u(\mf x)\hat{a}_e\ket{n}\!\!\bra{p}
        \nonumber\\&\:\:\:\:\:\:\:\:\:\:\:\:\:\:\:\:\:\:\:+e^{-\mathrm{i}(\Delta M - \omega_e)\tau}\bar{u}(\mf x)\slashed{j}(\mf x){\nu}_e\hat{a}_e^\dagger\ket{p}\!\!\bra{n}\Big).
    \end{align}
    Now, defining the ladder operators $\hat{\sigma}^+ = \hat{a}_e\ket{n}\!\!\bra{p}$ and $\hat{\sigma}^- = \hat{a}_e^\dagger\ket{p}\!\!\bra{n}$, we obtain a two-level system, whose excited state consists of a neutron and whose deexcited state consists of a proton and an electron. This two-level system is then localized by the spatial profile defined by the spinor $\slashed j(\mf x)u(\mf x)$ and couples linearly to the neutrino field $\nu(\mf x)$. A more detailed discussion of this reduction can be found in the works \cite{neutrinos,antiparticles}.
    
    In terms of the raising and lowering operators $\hat{\sigma}-$ and $\hat{\sigma}^+$, one obtains the following interaction Hamiltonian weight:
    \begin{align}\label{eq:neutrinoH}
        \hat{h}_I(\mf x) &= \frac{G_F}{\sqrt{2}}\left(e^{\mathrm{i}\Omega\tau}\bar{\nu}_e\Lambda^{\!(f)}\!(\mf x)\sigma^++e^{-\mathrm{i}\Omega\tau}\overline{\Lambda}^{\!(f)}\!(\mf x){\nu}_e\sigma^-\right),
    \end{align}
    where $\Omega = \Delta M - \omega_e$ is the effective energy gap and $\Lambda^{\!(f)}(\mf x) = \slashed{j}(\mf x) u(\mf x)$ is the spacetime smearing spinor, responsible for controlling both the spatial localization and time duration of the interaction. Effectively, we obtain a two-level fermionic particle detector model that interacts linearly with the neutrino field. For more details regarding this particle detector model and its symmetries, we refer the reader to the discussion in Subsection II B of \cite{antiparticles}.
    
    It is important to notice that in this model the ground and excited states couple to different sectors of the neutrino field: the deexcitation of this detector happens through the emission of an anti-particle or absorption of a particle, while the excitation of the detector happens through the emission of a particle, or absorption of an anti-particle. In terms of the effective two-level model, the excited state couples to the particle sector, while the ground state couples to the anti-particle sector. This is a similar feature to that of the linear complex scalar detector model presented in Section \ref{sec:anti}.

    \subsection{A particle detector probing a spin $1/2$ field}\label{sub:fermion}
    
    We consider a fermionic quantum field $\hat{\Psi}(\mf x)$ in a \mbox{$(3+1)$}-dimensional spacetime. We assume to have a basis of solutions of Dirac's equations split under the typical convention into so-called positive and negative frequency solutions, $\{u_{\bm p,s}(\mf x), v_{\bm p,s}(\mf x)\}$, so that the fermionic field can be written as
    \begin{align}\label{eq:fermionField}
        \hat{\Psi}(\mf x) &= \sum_{s=1}^2 \int \dd^3 \bm p\left( u_{\bm p,s}(\mf x)\hat{a}_{\bm p,s} + v_{\bm p,s}(\mf x) \hat{b}^\dagger_{\bm p,s}\right),\\
        \hat{\bar{\Psi}}(\mf x) &= \sum_{s=1}^2 \int \dd^3 \bm p\left( \bar{u}_{\bm p,s}(\mf x)\hat{a}^\dagger_{\bm p,s} + \bar{v}_{\bm p,s}(\mf x) \hat{b}_{\bm p,s}\right),
    \end{align}
    where $\hat{\bar{\Psi}}(\mf x)$ is the conjugate spinor to $\hat{\Psi}(\mf x)$ {and} $\hat{a}_{\bm p,s}$ and $\hat{b}_{\bm p,s}$ are the annihilation operators associated with particles and anti-particles, respectively. They satisfy the fermionic canonical anti-commutation relations
    \begin{align}
        \acomm{\hat{a}^{\vphantom{\dagger}}_{\bm p}}{\hat{a}^\dagger_{\bm p'} } &= \delta^{(3)}(\bm p - \bm p'),&&& \acomm{\hat{b}^{\vphantom{\dagger}}_{\bm p}}{\hat{b}^\dagger_{\bm p'} } &= \delta^{(3)}(\bm p - \bm p'),\nonumber\\
        \acomm{\hat{a}^{\vphantom{\dagger}}_{\bm p}}{\hat{a}^{\vphantom{\dagger}}_{\bm p'}\!\!\phantom{^\dagger} } &= 0,&&& \acomm{\hat{b}^{\vphantom{\dagger}}_{\bm p}}{\hat{b}^{\vphantom{\dagger}}_{\bm p'}\!\!\phantom{^\dagger} } &= 0,\label{car}\\
        \acomm{\hat{a}^\dagger_{\bm p}}{\hat{a}^\dagger_{\bm p'} } &= 0,&&& \acomm{\hat{b}^\dagger_{\bm p}}{\hat{b}^\dagger_{\bm p'} }&= 0.\nonumber
    \end{align}

    In order to generalize and simplify the linear fermionic particle detector presented in the previous subsection, we consider a particle detector with spinor spacetime smearing function $\Lambda^{\!(f)}\!(\mf x)$ coupled to a general spinor field $\hat{\Psi}(\mf x)$ via the following interaction Hamiltonian weight:
    \begin{equation}\label{hIfermion}
        \hat{h}_{I}(\mf x) = \lambda\left(e^{\mathrm{i}\Omega \tau}\hat{\sigma}^+\hat{\bar{\Psi}}(\mf x)\Lambda^{\!(f)}\!(\mf x) + e^{-\mathrm{i}\Omega \tau}\hat{\sigma}^-\bar{\Lambda}^{\!(f)}\!(\mf x)\hat{\Psi}(\mf x)\right).
    \end{equation}
    Notice that the Hamiltonian weight above is very similar to that of the linear complex particle detector. In fact, Eq. \eqref{hIfermion} follows from Eq. \eqref{hIcomplex} with the substitution
    \begin{equation}\label{hack}
       \hat{\psi}^\dagger(\mf x)\Lambda^{\!(c)}\!(\mf x)\longmapsto \hat{\bar{\Psi}}(\mf x)\Lambda^{\!(f)}\!(\mf x),
    \end{equation}
    where it should be understood that the spacetime smearing $\Lambda^{\!(c)}\!(\mf x)$, which is a complex function on the left-hand side, gets replaced by the spinor function $\Lambda^{\!(f)}\!(\mf x)$ on the right-hand side. We will use this fact to easily obtain analytical expressions for the linear fermionic detector.

    \section{Entanglement harvesting from a fermionic field with linearly coupled detectors }\label{sec:fermiHavest}
    
    We now turn our attention to the setup of entanglement harvesting. For this purpose, we consider two fermionic detectors undergoing trajectories $\mf z_j(\tau_j)$, \mbox{$j=1,2$}, with energy gaps $\Omega_j$, so that the interaction Hamiltonian weight associated with each of the two detectors is
    \begin{equation}\label{hIfermionj}
        \hat{h}_{I,j}(\mf x) \!=\! \lambda\!\left(\!e^{\mathrm{i}\Omega_j \tau_j}\hat{\sigma}_j^+\hat{\bar{\Psi}}(\mf x)\Lambda^{\!(f)}_j\!(\mf x) \!+\! e^{-\mathrm{i}\Omega_j \tau_j}\hat{\sigma}^-_j\overline{\Lambda}^{\!(f)}_j\!(\mf x)\hat{\Psi}(\mf x)\!\right)\!,
    \end{equation}
    where $\Lambda^{\!(f)}_j\!(\mf x)$ is the spacetime smearing spinor field associated with the $j$-th detector. \tb{Notice that in Eq. \eqref{hIfermionj} we have taken the coupling constants to be identical $\lambda_1 = \lambda_2 = \lambda$.} The total interaction Hamiltonian weight is therefore
    \begin{equation}\label{hIfermionsum}
        \hat{h}_{I}(\mf x) =  \hat{h}_{I,1}(\mf x)+\hat{h}_{I,2}(\mf x).
    \end{equation}
    
    Given that all analytical expressions for the fermionic particle detector can be obtained from the results of Section \ref{sec:antiHarvest} upon the replacement~\eqref{hack}, the differences between the linearly coupled scalar complex field harvesting and the fermionic case will come from the corresponding two-point functions. Same as before, due to the fact that in the vacuum state $\ket{0}$,
    \begin{align}
        \bra{0} \hat{\Psi}(\mf x) \hat{\Psi}(\mf x') \ket{0} &= 0,\\
        \bra{0} \hat{\bar{\Psi}}(\mf x) \hat{\bar{\Psi}}(\mf x') \ket{0} &= 0,
    \end{align}
    we see again that if the two detectors start in the ground state (or both in the excited state), they will be unable to harvest entanglement from the vacuum of the fermionic field. The same is true for any charge-conjugation invariant state. The reason for this is the same that we had for the linear coupled complex scalar fields: the anti-particle sector is only correlated with the particle sector.
    
    We are interested in harvesting entanglement, thus we will focus on the case where detector $1$ starts in the ground state, while detector $2$ starts in the excited state. In terms of the analogy with the physical model presented in Subsection \ref{sub:4fermi}, we would have one of the nucleons starting in the proton state, while the other starts in the neutron state. The density matrix associated with this process is then given by Eq. \eqref{rhoDge}, and its components are given by Eq. \eqref{Nijc}, where the term $\hat{\Psi}_j(\mf x)$ is given by    
    \begin{equation}
        \hat{\Psi}_j(\mf x) = \bar{\Lambda}^{\!(f)}_j\!(\mf x) \hat{\Psi}(\mf x).
    \end{equation}
    Notice that while the amount of entanglement that could be harvested in the complex scalar case was the same as the real scalar case, the fact that the Fermionic two-point functions are different imply noticeable changes in entanglement harvesting. In fact, the Wightman matrices for the fermionic field are given by 
    \begin{align}
        W_0(\mf x,\mf x') &= \bra{0}{\hat{\Psi}}(\mf x)\hat{\bar{\Psi}}(\mf x')\ket{0} =\sum_{s=1}^2\int \dd^3 \bm p\: u_{\bm p,s}(\mf x)\bar{u}_{\bm p,s}(\mf x'),\nonumber\\
        \overline{W}_0(\mf x,\mf x') &= \bra{0}\hat{\bar{\Psi}}(\mf x)\hat{\Psi}(\mf x')\ket{0} = \sum_{s=1}^2\int \dd^3 \bm p\: v_{\bm p,s}(\mf x')\bar{v}_{\bm p,s}(\mf x) \label{eq:Woconjugate}.
    \end{align}
    
    In order to explicitly compare the difference between the entanglement harvested from complex scalar fields and fermionic fields, we consider an explicit example in flat spacetime. We choose the following basis of solutions to Dirac's equations in Minkowski spacetime:
    \begin{align}
        u_{\bm p,1}(\mf x)&=\frac{1}{(2\pi)^{\frac{3}{2}}}\sqrt{\frac{\omega_{\bm p}+m }{2 \omega_{\bm p} }}\left(\begin{array}{c}
        1 \\
        0 \\
        \frac{p_{z} }{\omega_{\bm p}+m } \\
        \frac{p_{x}+\mathrm{i} p_{y}}{\omega_{\bm p}+m }
        \end{array}\right)e^{\mathrm{i} \mf p\cdot \mf x},\\
        u_{\bm p,2}(\mf x)&=\frac{1}{(2\pi)^{\frac{3}{2}}}\sqrt{\frac{\omega_{\bm p}+m }{2 \omega_{\bm p} }}\left(\begin{array}{c}
        0\\
        1\\
        \frac{p_{x}-\mathrm{i} p_{y}}{\omega_{\bm p}+m }\\
        \frac{-p_z}{\omega_{\bm p}+m }  
        \end{array}\right)e^{\mathrm{i} \mf p\cdot \mf x},
        \\
        v_{\bm p,1}(\mf x)&=\frac{1}{(2\pi)^{\frac{3}{2}}}\sqrt{\frac{\omega_{\bm p}+m }{2 \omega_{\bm p} }}\left(\begin{array}{c}
        \frac{p_{z} }{\omega_{\bm p}+m } \\
        \frac{p_{x}+\mathrm{i} p_{y} }{\omega_{\bm  p}+m } \\
        1 \\
        0
        \end{array}\right)e^{-\mathrm{i} \mf p\cdot \mf x},\\\textcolor{white}{Vanier}\nonumber
        \end{align}\begin{align}%
        v_{\bm p,2}(\mf x) &=\frac{1}{(2\pi)^{\frac{3}{2}}}\sqrt{\frac{\omega_{\bm p}+m }{2 \omega_{\bm p} }}\left(\begin{array}{c}
        \frac{p_{x}-\mathrm{i} p_{y} }{\omega_{\bm p}+m } \\
        \frac{-p_{z} }{\omega_{\bm p}+m } \\
        0 \\
        1
        \end{array}\right)e^{-\mathrm{i} \mf p\cdot \mf x}.\vspace{5mm}
    \end{align}
    This basis is normalized so that the canonical anti-commutation relations from Eqs. \eqref{car} hold. These also satisfy the completeness relations
    \begin{align}
        \sum_{s=1}^2 u_{\bm p,s}(\mf x) u_{\bm p,s}(\mf x') &= \frac{1}{(2\pi)^3} \frac{e^{\mathrm{i} \mf p \cdot ( \mf x-\mf x')}}{2\omega_{\bm p}}(\slashed p + m),\\
        \sum_{s=1}^2 v_{\bm p,s}(\mf x') v_{\bm p,s}(\mf x) &= \frac{1}{(2\pi)^3} \frac{e^{\mathrm{i} \mf p \cdot( \mf x-\mf x')}}{2\omega_{\bm p}}(\slashed p - m).
    \end{align}
    
    We pick the spacetime spinor smearing functions to be given by
    \begin{align}
        \Lambda^{\!(f)}_1(\mf x) &= \chi(t)F(\bm x)\eta_1,\\
        \Lambda^{\!(f)}_2(\mf x) &= \chi(t)F(\bm x-\bm L)\eta_2,
    \end{align}
    where $\chi(t)$ and $F(\bm x)$ are given by Eqs. \eqref{chi} and \eqref{F} so that $\sigma$ is the average size of the interaction region and $T$ is its average time duration. $\eta_1$ and $\eta_2$ are arbitrary constant spinors.
    
    We now focus on the interaction of the detectors with the vacuum of the fermionic field. We obtain the following density matrix for the detectors after the interaction with the vacuum:
    
     \begin{equation}
        \hat{\rho}_{\text{d}} = \begin{pmatrix}
            {\tb{\mathcal{L}_{11}'}} & 0 & 0 & 0\\
            0 & 0 & \mathcal{M}' & 0 \\
            0 & \mathcal{M}'^* & 1 - {\tb{\mathcal{L}_{11}'}} - \mathcal{L}_{22} & 0\\
            0 & 0 & 0 & \mathcal{L}_{22}
        \end{pmatrix}+ \mathcal{O}(\lambda^3).
    \end{equation}
    We can lighten the notation by noticing that \mbox{$\tb{\L_{11}'} = \L(-\Omega)$} and ${\L}_{22} = \L(\Omega)$, where
    
    \begin{widetext}
    \begin{equation}
    \begin{gathered}
        {\mathcal{L}(\Omega)} = \frac{\lambda^2}{(2\pi)^3}\int \frac{\dd^3 \bm p}{2 \omega_{\bm p}}{\color{black}\bar{\eta}_i(\slashed p + m)\eta_i}|\tilde{F}(\bm p)|^2 |\tilde{\chi}( \omega_{\bm p}+\Omega)|^2,\\
        \mathcal{M}' = \frac{\lambda^2}{(2\pi)^3} \int \frac{\dd^3 \bm p}{2 \omega_{\bm p}} |\tilde{F}(\bm p)|^2 e^{\mathrm{i} \bm p \cdot \bm L}({\color{black}\bar{\eta}_1(\slashed p + m)\eta_2}Q(-\omega_{\bm p}- \Omega) {\color{black}+}{\color{black}\bar{\eta}_1(\slashed p - m)\eta_2}Q(\Omega-\omega_{\bm p}) ).
    \end{gathered}\label{mlf}
    \end{equation}
    \end{widetext}
    The function $Q(\omega)$ is given by Eq. \eqref{Comega}, and the tilde denotes Fourier transform, as in Section \ref{sec:antiHarvest}.
    

    We can now study the differences between harvesting from complex scalar fields and from fermionic fields with linear couplings. The major difference is related to the \tb{factors of} \mbox{$\bar{\eta}_i ( \slashed p \pm  m)\eta_j$}, which show up in the integrals above. A consequence of this is the fact that the detectors can, in principle, have different transition probabilities depending on the choices of $\eta_1$ and $\eta_2$. The last noticeable difference is the units of the coupling constant in this model, which was dimensionless in Section \ref{sec:anti}, but has dimension of $[E]^{-2}$ here.
    
    In order to compute explicit examples, we must choose the spinors $\eta_1$ and $\eta_2$. 
   We make the following choice:
    \begin{equation}\label{Lambda12Eq}
        \eta_1 = \eta_2 =\begin{pmatrix}
            \Delta\\0\\0\\0
        \end{pmatrix}.
    \end{equation}
    In the case of the interaction of nucleons with the neutrino fields, $\Delta$ would be associated with the energy scale of the electrons involved in the interaction. In essence, the choices of $\eta_1$ and $\eta_2$ are associated with the electron states involved in the interaction with the neutrino field. The assumption that $\eta_1 = \eta_2$ above then indicates that the electron states involved in both processes are the same. This choice should generally increase the correlations between the detectors by making them associated to processes that involve electrons with the same energy and spin. In fact, Eq. \eqref{Lambda12Eq} picks both electron states to have the spin in the positive $z$ direction in the chosen coordinate system. 
    
    With the choices from Eq. \eqref{Lambda12Eq}, we obtain
    \begin{equation}
        \bar{\eta}_1 \slashed{p} \eta_2 = \Delta^2\, \omega_{\bm p} = \Delta^2 \sqrt{\bm p^2 + m^2}, \quad\quad \bar{\eta}_1  m \eta_2 = \Delta^2 m,
    \end{equation}
    where $\Delta$ is a constant with dimensions of $[E]^\frac{3}{2}$ that takes into account the units of $\eta_1, \eta_2$. The results above allow one to solve Eq. \eqref{mlf} analytically in the angular variables, so that once more we are left with one integral with respect to $|\bm p|$ that can be done numerically.

    \begin{figure}[h]
        \includegraphics[scale=0.34]{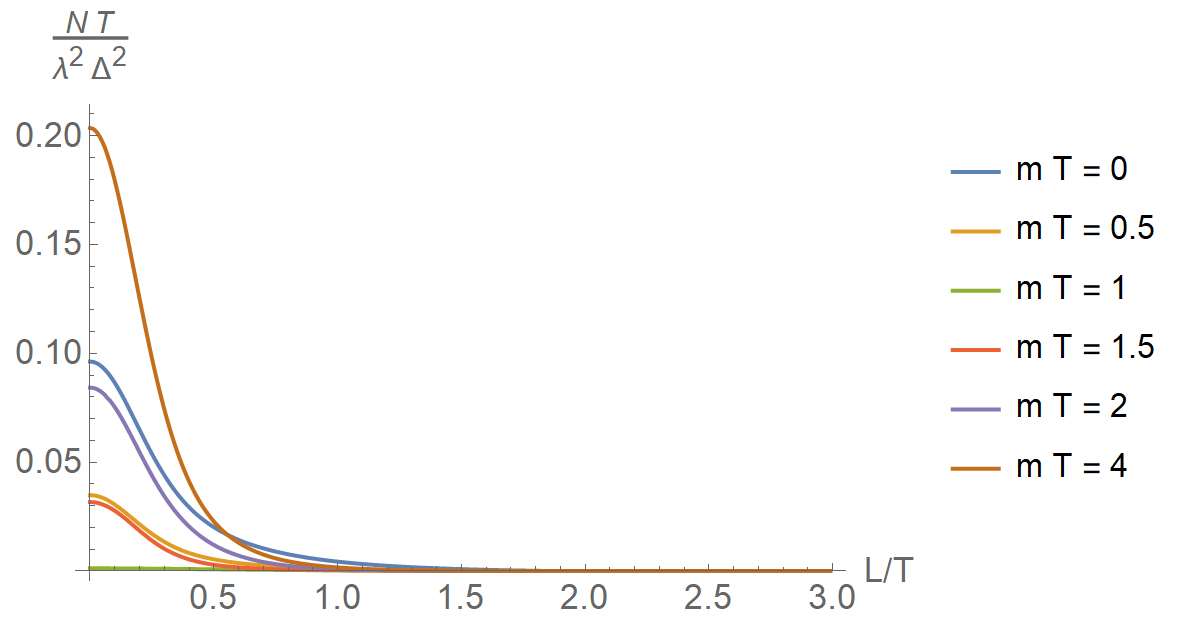}
        \caption{\trr{Negativity of the two detector system as a function of the detectors separation for different values of the field mass. We fixed the detector gap as $\Omega T = 1$ and the detector size as $\sigma = 0.1 T$. This is analogous to the plot in Fig. \ref{fig:scalar}.  }}\label{fig:fermionic}
    \end{figure}

    \begin{figure}[h]
        \includegraphics[scale=0.34]{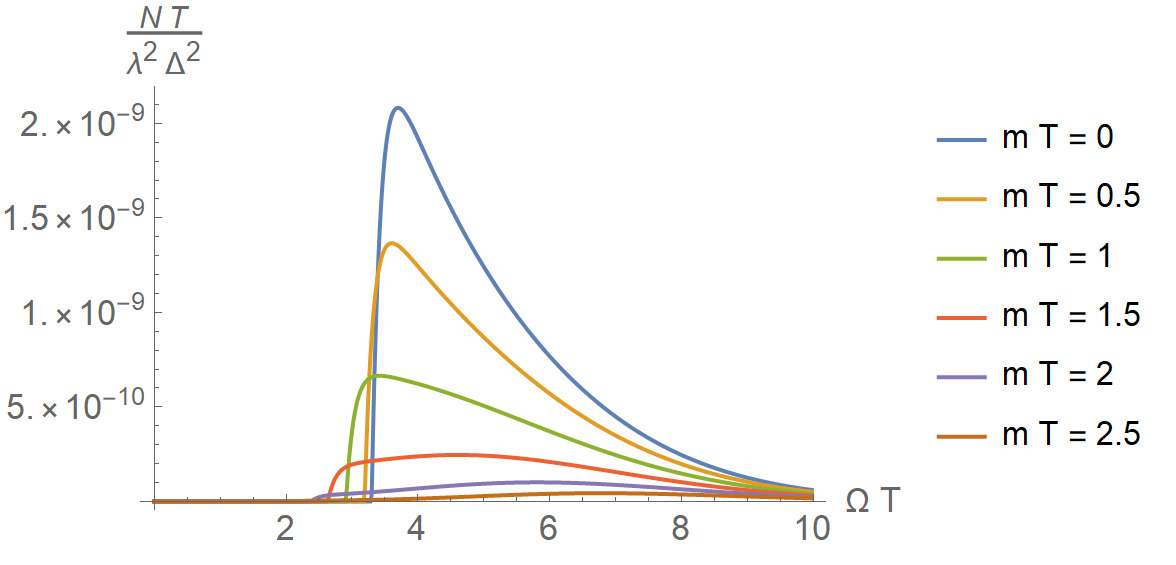}
        \caption{\trr{Negativity of the two detector system as a function of the detector gap for different values of the field mass. We fixed the detector separation as $L = 5 T$ and the detector size as $\sigma = 0.2 T$. This is analogous to the plot in Fig. \ref{fig:scalar1}. }}\label{fig:fermionic1} 
    \end{figure}
    
        \begin{figure}[h]
        \includegraphics[scale=0.34]{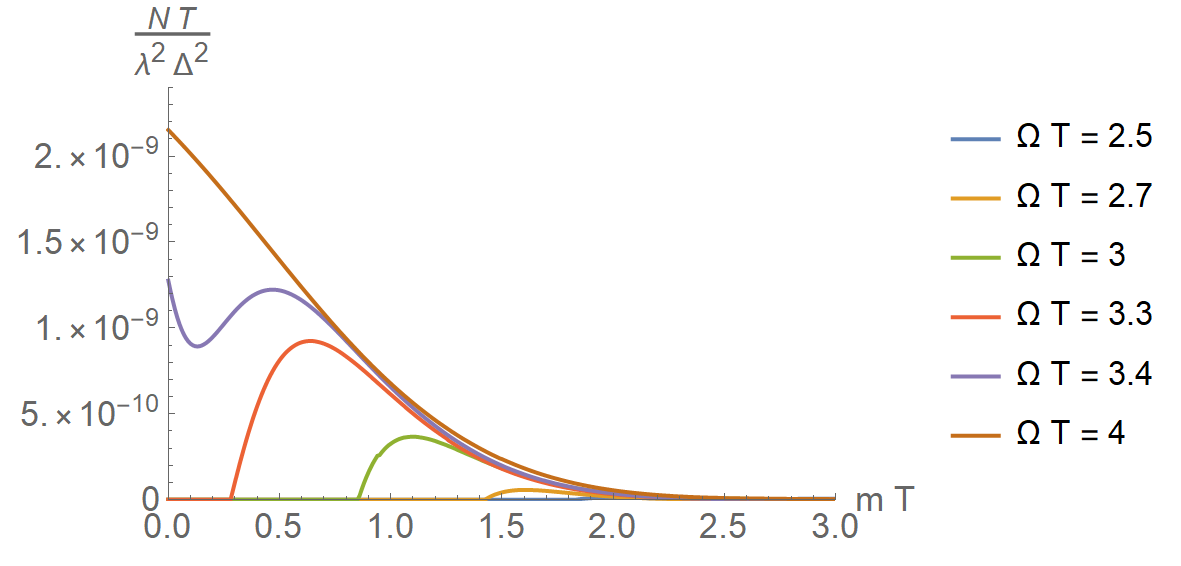}
        \caption{\trr{Negativity of the two detector system as a function of the field's mass for different values of $\Omega$. We fixed the detector separation as $L = 5 T$ and the detector size as $\sigma = 0.1 T$. These are analogous to the plots in Fig. \ref{fig:scalar2}.}}\label{fig:fermionic2} 
    \end{figure}
    
    \begin{figure}[h]
        \includegraphics[scale=0.34]{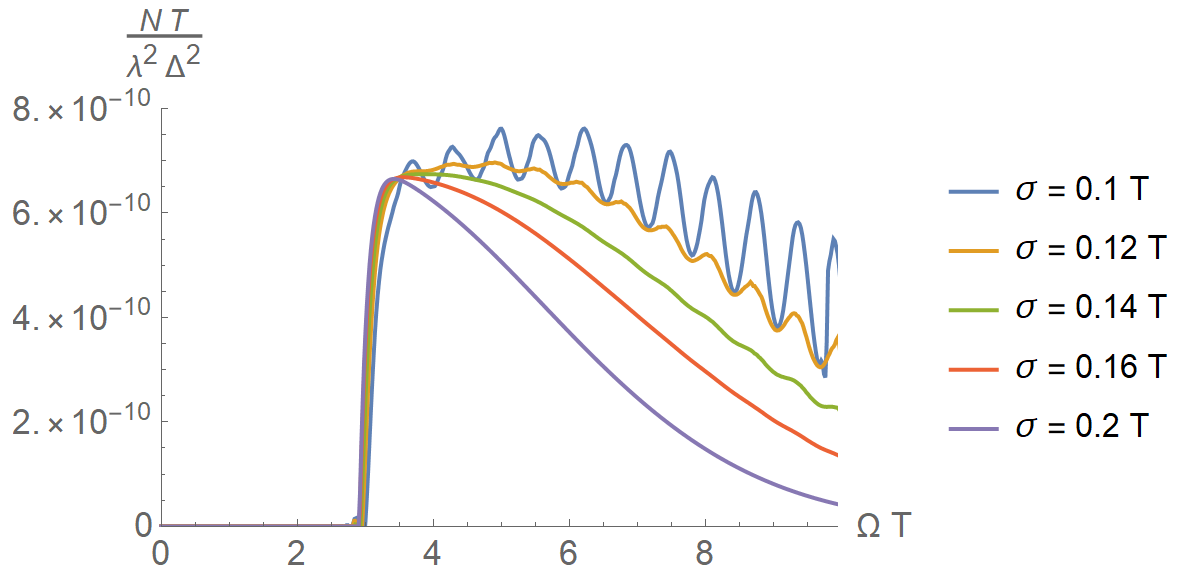}
        \caption{\trr{Negativity of the two detector system as a function of the detector gap $\Omega$ for different values of $\sigma$. We fixed the detector separation as $L = 5 T$ and the \tbb{field's mass so that} \mbox{$m T = 1$}. This is analogous to the plot in Fig. \ref{fig:scalarOsc}. }}\label{fig:fermionicOsc}
    \end{figure}

    \begin{figure}[h!]
        \includegraphics[scale=0.38]{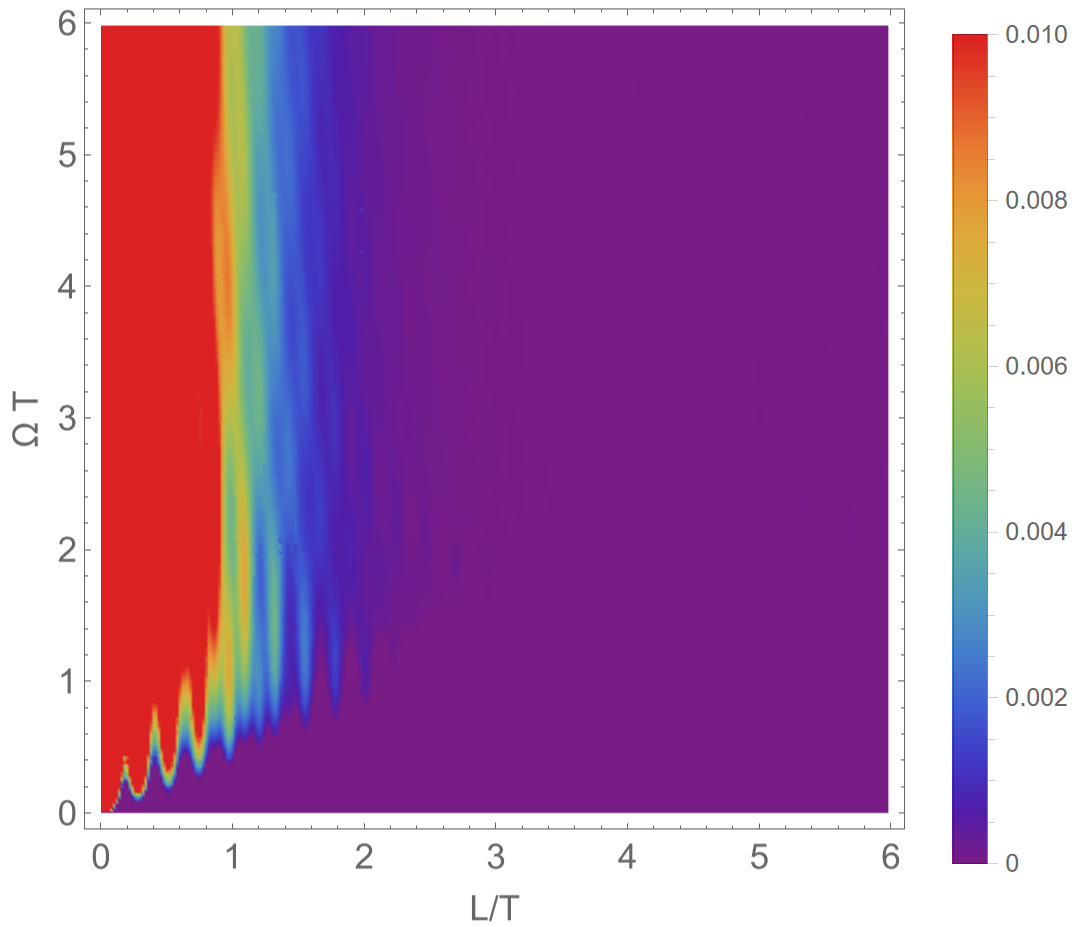}
        \caption{\trr{Negativity of the two detector system as a function of the detectors gap and separation. We fixed $m=0$ and considered pointlike detectors with $\sigma = 0$. This is analogous to the plot in Fig. \ref{fig:scalar2D}.}}\label{fig:fermionic2D}
    \end{figure}
    
    \begin{figure}[h!]
        \includegraphics[scale=0.4]{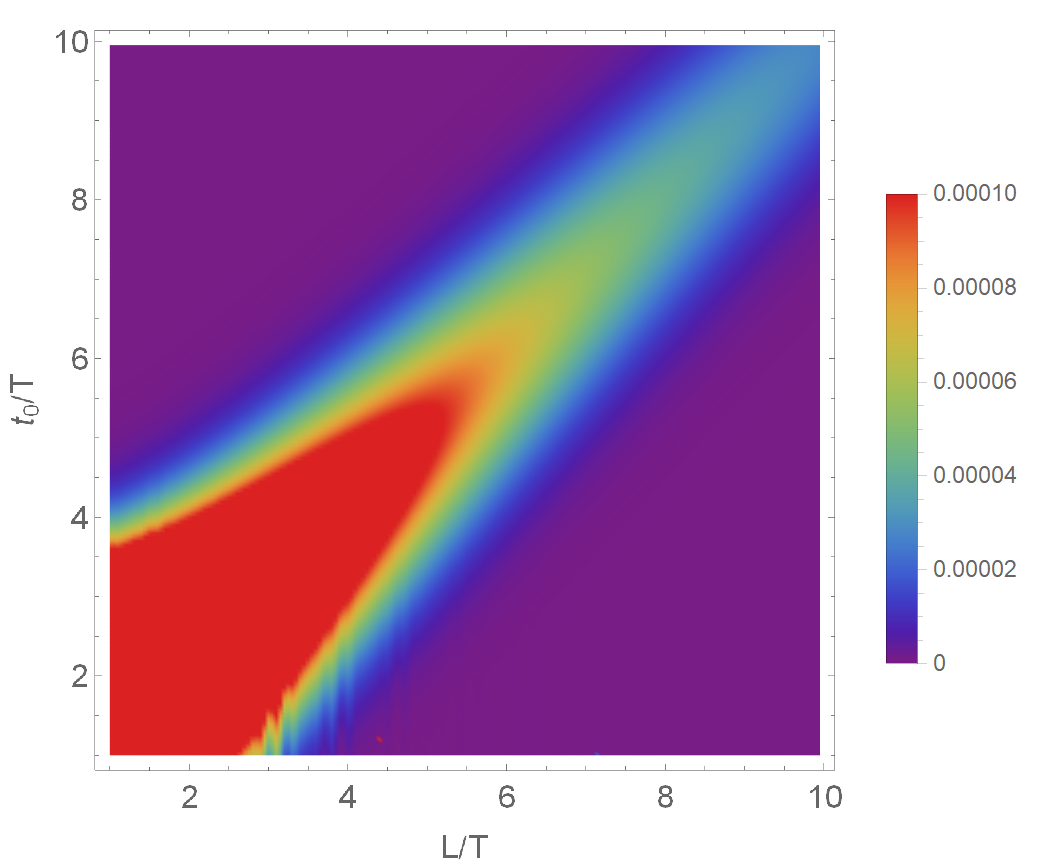}
        \caption{Negativity of the two detector system as a function of the detectors separation in space ($L$) and in time ($t_0$). We fixed $\Omega = 4/T$ for the detector gap, $m=0$ for the field mass, and considered pointlike detectors with $\sigma = 0$. This is analogous to the plot in Fig. \ref{fig:scalar2DSpacetime}. }\label{fig:fermionic2DSpacetime}
    \end{figure}
    
    {\color{black} In Figs. \ref{fig:fermionic}, \ref{fig:fermionic1}, \ref{fig:fermionic2}, \ref{fig:fermionicOsc}, \ref{fig:fermionic2D}, \ref{fig:fermionic2DSpacetime}, \ref{fig:azLf}, \ref{fig:azOmegaf}, \ref{fig:axLf}, and \ref{fig:axOmegaf} we plot the negativity of the detectors state as a function of the different parameters of the setup. We choose \tb{similar} parameters that we did in the case of the scalar model in Subsection \ref{sub:UDWharvestge}, such that we can establish a direct comparison between the pairs of Figures
    \ref{fig:fermionic} and \ref{fig:scalar}, \ref{fig:fermionic1} and \ref{fig:scalar1},
    \ref{fig:fermionic2} and \ref{fig:scalar2},
    \ref{fig:fermionicOsc}  and \ref{fig:scalarOsc},
    \ref{fig:fermionic2D} and \ref{fig:scalar2D},  \ref{fig:fermionic2DSpacetime} and \ref{fig:scalar2DSpacetime}, \ref{fig:azLf} and \ref{fig:azL}, \ref{fig:azOmegaf} and \ref{fig:azOmega}, \ref{fig:axLf} and \ref{fig:axL}, and \ref{fig:axOmegaf} and \ref{fig:axOmega}. \trr{In general, we see a very similar behaviour of the negativity in the fermionic and complex scalar cases, with minor differences that we point out below.}
    
    In Fig. \ref{fig:fermionic} we plot the negativity as a function of the detectors separation for different masses of the field. Overall, we see that the negativity decays rapidly with $L$, same as in the scalar case. \trr{However, here we see that for small values of $L$, the negativity oscillates rapidly with the field's mass, unlike in the complex scalar case. Nevertheless, we note that the case of \tbb{$L<\sigma$} is unphysical, as the detectors are superposed in that regime. Analogously, for values of $L$ smaller than $5T$ the spacetime smearing functions are considerably timelike separated, so that this does not configure entanglement harvesting.}
        
    In Figs. \ref{fig:fermionic1} and \ref{fig:fermionic2} we study the behaviour of the negativity of the detectors' state as a function of their detectors gaps and the field mass when the detectors are separated by a distance $L = 5T$, which makes the effective interaction regions of the detectors spacelike separated \tb{(}with more than 10$\sigma$ of separation\tb{)}. In Fig. \ref{fig:fermionic1} the negativity is plotted as a function of the detectors gap for varying values of the field mass. \tb{Unlike in the complex scalar case, we see a monotonous behaviour for negativity with mass, with peaks decreasing as mass increases.} \tb{On the other hand, same as in the scalar case, mass decreases the amount of entanglement that can be harvested, but allows one to harvest using detectors with smaller energy gaps.} In Fig. \ref{fig:fermionic2}, the negativity of the detectors state is plotted as a function of the field's mass for different detectors energy gaps. \tb{We also see a similar behaviour in the fermionic case and the complex one, provided that the field's mass is larger than the approximate value of $mT > 0.25$.} 
    
    In Fig. \ref{fig:fermionicOsc} we plot the analogue of Fig. \ref{fig:scalarOsc}, that is, we study the behaviour of the negativity of the detectors state for varying detector gaps and detector sizes. 
    \trr{Similar to the complex scalar case, we see oscillations for small values of $\sigma$. We also remark that the amplitude of the oscillations is suppressed by the detector's size.} 

    Figure \ref{fig:fermionic2D} shows the negativity as a function of $L$ and $\Omega$. Comparing this plot to Fig. \ref{fig:scalar2D}, \tb{we find an overall similar behaviour, with minor differences. } \trr{We see here that the main differences between the complex and the fermionic plots as a function of $\Omega$ and $L$ show up for small $L$, where the detectors have considerable overlap. In these regimes, one could claim that the fermionic detector model is not well approximated by the complex scalar model, but these are also the regimes which are less physical and interesting for the protocol of entanglement harvesting.} \tb{We can also see the oscillations present in Fig. \ref{fig:fermionicOsc} in Fig. \ref{fig:fermionic2D}.} Finally, in Fig. \ref{fig:fermionic2DSpacetime} we consider the situation where the centers of the detectors' interactions are separated in time by $t_0$ and in space by $L$. Same as in the scalar case, we observe that acquired entanglement peaks near the lightcone, although the peak most likely corresponds to entanglement acquired through communication rather than harvesting~\cite{ericksonNew}.

    \tb{Finally, we consider complex smearing functions, same as we did in the complex scalar case, by adding a factor of $e^{\ii \bm a \cdot \bm x}$ to the smearing functions of both detectors. Using expressions in Eq. \eqref{mlf}, we can then compute the negativity associated to both detectors. Choosing the $z$ axis parallel to the separation between the detectors, $\bm L$, we plot the negativity as a function of the components of the vector $\bm a$ in Figs. \ref{fig:azLf}, \ref{fig:azOmegaf}, \ref{fig:axLf} and \ref{fig:axOmegaf}}
    
    \tb{In Figs. \ref{fig:azLf} and \ref{fig:azOmegaf} we plot the negativity as a function of the component of $\bm a$ parallel to $\bm L$ for varying values of $\bm L$ and $\Omega$. Same as in the complex scalar case, we see that it is possible to increase the negativity by two orders of magnitude by choosing $a_z \approx \Omega$. In Fig. \ref{fig:azLf} we see the expected decay of negativity with the detectors separation. In Fig. \ref{fig:azOmegaf} we can clearly see the peaks of negativity when $a_z \approx \Omega$. Once again we see that by regulating the complex phase of the smearing function it is possible to enhance the entanglement harvested from a fermionic field.}
    
    \tb{In Figs. \ref{fig:axLf} and \ref{fig:axOmegaf} we plot the negativity of the detectors system as a function of the orthogonal component of $\bm a$ to $\bm L$, $a_x$. In both plots we see that the negativity only decreases as the orthogonal component of $\bm a$ to the detectors separation increases. The overall behaviour of these plots is very similar to the one seen in the complex scalar case, in Figs. \ref{fig:axL} and \ref{fig:axOmega}.}
    
    \begin{figure}[h]
        \includegraphics[scale=0.36]{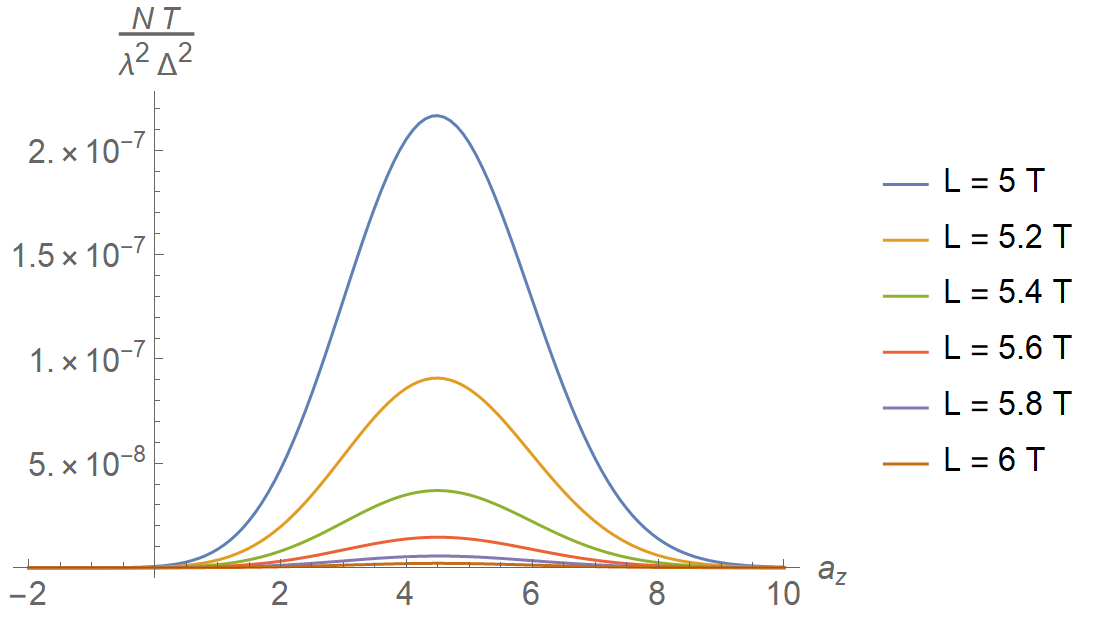}
        \caption{\tb{Negativity of the two detector system as a function of the parallel component of the vector parameter $\bm a$ for varying $L$ with $a_x = 0$. We fixed $m T =0.5$ and considered $\sigma = 0.5 T$. }}\label{fig:azLf}
    \end{figure}
    
    \begin{figure}[h]
        \includegraphics[scale=0.36]{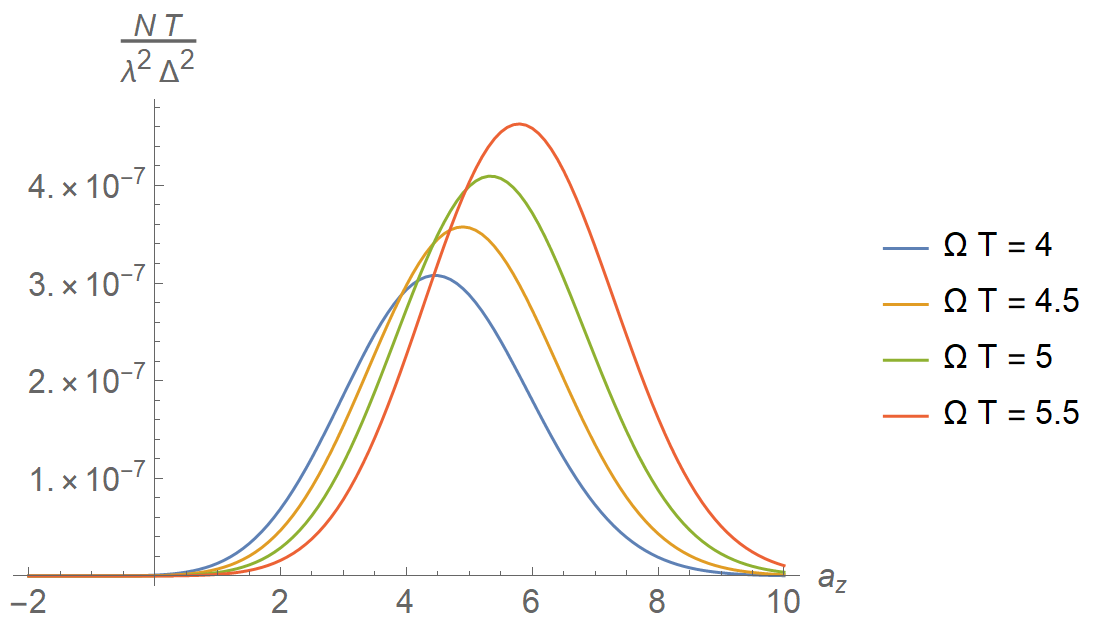}
        \caption{\tb{Negativity of the two detector system as a function of the parallel component of the vector parameter $\bm a$ for varying $\Omega$ with $a_x = 0$. We fixed $m T =0.5$ and considered $\sigma = 0.5 T$.  }}\label{fig:azOmegaf}
    \end{figure}
    
    \begin{figure}[h]
        \includegraphics[scale=0.36]{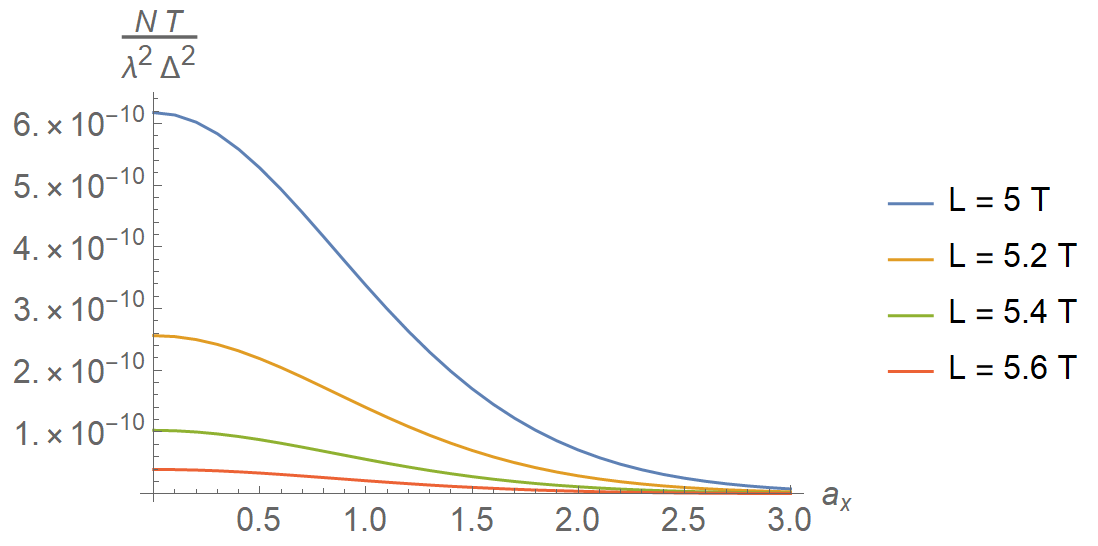}
        \caption{\tb{Negativity of the two detector system as a function of the orthogonal component of the vector parameter $\bm a$ for varying $L$ with $a_z = 0$. We fixed $m T =0.5$ and considered $\sigma = 0.5 T$. }}\label{fig:axLf}
    \end{figure}
    
    \begin{figure}[h!]
        \includegraphics[scale=0.36]{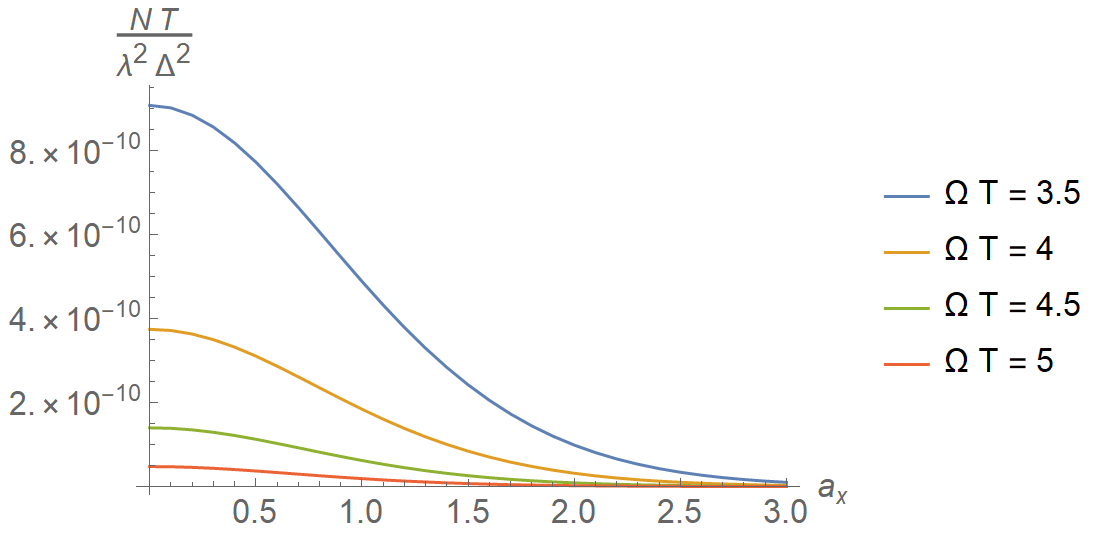}
        \caption{\tb{Negativity of the two detector system as a function of the orthogonal component of the vector parameter $\bm a$ for varying $\Omega$ with $a_z = 0$. We fixed $m T =0.5$ and considered $\sigma = 0.5 T$.}}\label{fig:axOmegaf}
    \end{figure}
    
    \trr{Overall, we find that the behaviour of the negativity for the fermionic detectors is very similar to the complex scalar case, with mild differences only in the case where the detector separation is small and when the mass of the field compared to the inverse time duration of the interaction is small. In all other regimes, we find that the complex scalar model can be used as a good approximation for the entanglement harvesting protocol with the linear fermionic model studied in this section.}
    
    
    \tb{As mentioned before, the neutron decay into a proton, electron and anti-neutrino (Eq. \ref{betadecay}) can be modelled by this fermionic particle detector. We can then apply our above calculations to estimate the amount of entanglement that can be harvested from two of those detectors. Considering realistic values for $\Omega, \lambda, \Delta$ \tbb{(\mbox{$\Omega \approx m_n \!- \!m_p\!-\!me\approx\!  0.78 \text{MeV}$}, $\lambda\approx\! 0.82\times 10^{-5} \text{GeV}^{-2}$,\vspace{-0.5mm} $\Delta\approx m_e^{3/2} \approx 0.51 \text{MeV}$)}, taking $\Omega T=1$ and noticing that the peaks of negativity we obtained are of order $10^{-9} \lambda^2 \Delta^2 T^{-1}$ (Figs. \ref{fig:fermionic1}, \ref{fig:fermionic2}, and \ref{fig:fermionicOsc}), we obtain a negativity of the order $10^{\tbb{-33}}$. Comparing this result with the one obtained for the case of a hydrogenlike atom harvesting entanglement from the electromagnetic field, studied in \cite{Pozas2016}, where the negativity found was of the order of $10^{-46}$ \tbb{(Fig. 6 of~\cite{Pozas2016})}, we see that the negativity harvested in the beta decay is considerably larger. Furthermore, using a complex smearing function, we were able to increase the harvested negativity by two orders of magnitude (Figs. \ref{fig:azLf}, \ref{fig:azOmegaf}), i.e., $10^{-39}$.}
     
    }
    
        

\section{Conclusions}\label{sec:conclusion}
    
    \vspace{-1mm}
    
    We have \tb{performed a comparative study of} entanglement harvesting with particle detector models that couple linearly to \tb{real and} complex scalar and fermionic fields. \tb{While the fermionic particle detector model is inspired by realistic high-energy physics processes~\cite{neutrinos,antiparticles}, the complex detector has been shown to be a good approximation for the fermionic one when dealing with particle states~\cite{antiparticles}}. By using linear models we can avoid the persistent divergences that appear when studying entanglement harvesting with quadratic couplings~\cite{Sachs1,sachs2018entanglement,eduAchimBosonFermion}. We have formally computed the density matrix of {pairs} of detectors moving on arbitrary trajectories in curved spacetimes when the field is in a general state. Then, we explicitly solved examples of comoving inertial detectors in Minkowski spacetime. In that scenario we quantified the amount of entanglement acquired by two linearly coupled detectors through the negativity of the detectors state after the interaction with the field. This allowed us to compare the \tb{entanglement harvesting protocol for the} fermionic and complex scalar models. \tb{Overall, we found that these models behave very similarly.}
    
    For \tb{both} complex scalar and fermionic detector models, we have discussed how the ground state only couples to the anti-particle sector of the field. Hence, two detectors that both start in the ground or excited state are unable to harvest entanglement from the vacuum state of a complex/fermionic quantum field. This result holds independently of the detectors trajectories, shape of the interaction or quantization scheme. We showed that this originates from the fact that all the correlations of non-Hermitian fields manifest through correlations of the particle and anti-particle sectors. 
    
    On the other hand, if one detector starts in the ground state, while the other starts in the excited state, we showed that linearly coupled complex as well as fermionic particle detectors can harvest entanglement from the vacuum of a quantum field. Comparing with the known results for real scalar fields, we showed that two complex scalar detector models can harvest the same amount of entanglement as two UDW detectors coupled to a real scalar field \tb{provided we consider the smearing function to be real. Moreover, we showed that by adding a complex phase term to the detector's smearing function, we can increase the harvested entanglement by up to two orders of magnitude.} 
    
     \trr{Another major result of this manuscript is that the complex scalar model grasps most of the features of the fermionic model when the mass of the fields is sufficiently large. This confirms the results of~\cite{antiparticles}, showing that the linearly coupled complex particle detector model is a good approximate description for the linear fermionic model even when two detectors are considered. In the same fashion that the massless real scalar UDW model can be used to describe the light-matter interaction when angular momentum is neglected, we argue that the linear complex scalar model can be used to model the interaction of nucleons with the neutrino fields when their spin does not play a major role.}
    
    

    \vspace{-1mm}
    
    \section{Acknowledgements}
    
    \vspace{-2mm}
    
    {The authors thank H\'ector Maeso-Garc\'ia for his helpful feedback on the first draft of the manuscript \tbb{and an anonymous referee for insightful suggestions.}} Research at Perimeter Institute is supported in part by the Government of Canada through the Department of Innovation, Science and Industry Canada and by the Province of Ontario through the Ministry of Colleges and Universities. E. M-M. is funded by the NSERC Discovery program as well as his Ontario Early Researcher Award.

\appendix

    \onecolumngrid

\section{Computations regarding the Ground - Ground Interaction of the Linear Complex Detector}\label{app:gg}

    Here we present the details of the computations regarding the general entanglement harvesting setup described in \tb{Subsection \ref{sub:antiHarvestgg}. We remind the reader that the complex scalar particle detector models reduce to the real scalar case if $\hat{\psi}(\mf x) = \hat{\psi}^\dagger(\mf x)$ and the spacetime smearing function is chosen as a real function. Noting this, we see that the result of this appendix can also be applied for the setup of entanglement harvesting of Subsection \ref{sub:UDWharvest}, considering $\hat{\psi}(\mf x) = \hat{\phi}(\mf x)$.} For the ground-ground initial state, the initial state density operator is
    \begin{equation}
        \hat{\rho}_0 = \ket{g_1}\!\!\bra{g_1}\otimes\ket{g_2}\!\!\bra{g_2}\otimes \hat{\rho}_{\phi} = \hat{\rho}_{\text{d}, 0}\otimes \hat{\rho}_{\phi}.
    \end{equation}
    Using the second order Dyson expansion for the time evolution operator in Eq. \eqref{itstnow}, we obtain
    \begin{equation}
        \hat{U}_I = \mathcal{T}\exp\left(-\mathrm{i} \int \dd V \hat{h}_I(\mf x)\right) = \openone +  \hat{U}_I^{(1)} + \hat{U}_I^{(2)}+\mathcal{O}(\lambda^3),
    \end{equation}
    where
    \begin{align}
        \hat{U}_I^{(1)}&=-\mathrm{i} \int \dd V \hat{h}_I(\mf x),\\
        \hat{U}_I^{(2)}&=- \int \dd V \dd V'\hat{h}_I(\mf x)\hat{h}_I(\mf x')\theta(t-t'),
    \end{align}
    and
    \begin{equation}
         \hat{h}_I(\mf x) = \lambda\Big(\left(\Lambda_1(\mf x)e^{\mathrm{i}\Omega_1 \tau_1}\hat{\sigma}_1^+\hat{\psi}^\dagger(\mf x) + \Lambda^*_1(\mf x)e^{-\mathrm{i}\Omega_1 \tau_1}\hat{\sigma}_1^-\hat{\psi}(\mf x)\right) +\left(\Lambda_2(\mf x)e^{\mathrm{i}\Omega_2 \tau_2}\hat{\sigma}_2^+\hat{\psi}^\dagger(\mf x) + \Lambda^*_2(\mf x)e^{-\mathrm{i}\Omega_2 \tau_2}\hat{\sigma}_2^-\hat{\psi}(\mf x)\right)\Big).
    \end{equation}
    The time evolved density operator for the full system consisting of the two-detectors and the field then reads
    \begin{equation}
        \hat{\rho} = \hat{U}_I\hat{\rho}_0\hat{U}_I^\dagger.
    \end{equation}
    We then find that, up to second order in $\lambda$, the detector density operator is given by
    \begin{align}
     \hat{\rho}_{\text{d}} = \tr_\phi(\hat{\rho}) =&\hat{\rho}_{\text{d},0}+\hat{\rho}^{(1)}_{\text{d}}+\hat{\rho}^{(1,1)}_{\text{d}}+\hat{\rho}^{(2,0)}_{\text{d}}+\hat{\rho}^{(0,2)}_{\text{d}} + \tbb{\mathcal{O}(\lambda^3)},
    \end{align}
    where
    \begin{align}
       \hat{\rho}_{\text{d}}^{(1)} &= \tr_\phi\left(U_I^{(1)}\hat{\rho}_0+\hat{\rho}_0U_I^{(1)\dagger}\right),\\ \hat{\rho}^{(1,1)}_{\text{d}} &=\tr_\phi\left(U_I^{(1)}\hat{\rho}_0U_I^{(1)\dagger}\right)= \int \dd V\dd V'\ev{\hat{h}_I(\mf x)\hat{\rho}_{\text{d}, 0}\hat{h}_I(\mf x')}_{\rho_{\phi}},\label{rho11}\\
        \hat{\rho}^{(2,0)}_{\text{d}} &=\tr_\phi\left(U_I^{(2)}\hat{\rho}_0\right) =  -\int \dd V\dd V'\theta(t-t')\ev{\hat{h}_I(\mf x)\hat{h}_I(\mf x')}_{\rho_{\phi}}\!\!\!\hat{\rho}_{\text{d}, 0} ,\\
        \hat{\rho}^{(0,2)}_{\text{d}} &=\tr_\phi\left(\hat{\rho}_0U_I^{(2)\dagger}\right)= -\int \dd V\dd V'\theta(t-t')\hat{\rho}_{\text{d}, 0}\!\ev{\hat{h}_I(\mf x)\hat{h}_I(\mf x')}_{\rho_{\phi}}.
    \end{align}
    Let us compute each of the above terms separately. The first order terms in the first parenthesis yields
    \begin{align}
        \hat{\rho}_{\text{d}}^{(1)} = -\mathrm{i}\lambda \int \dd V\Bigg(& \Lambda_1(\mf x )e^{\mathrm{i}\Omega_1 \tau_1}\langle\hat{\psi}^\dagger(\mf x) \rangle_{{\rho}_{\phi}}\ket{e_1g_2}\!\!\bra{g_1g_2}+\Lambda_2(\mf x)e^{\mathrm{i}\Omega_2 \tau_2}\langle\hat{\psi}^\dagger(\mf x) \rangle_{{\rho}_{\phi}}\ket{g_1e_2}\!\!\bra{g_1g_2}\nonumber\\
        &-\Lambda_1^*(\mf x)e^{-\mathrm{i}\Omega_1 \tau_1}\langle\hat{\psi}(\mf x) \rangle_{{\rho}_{\phi}}\ket{g_1g_2}\!\!\bra{e_1g_2}-\Lambda_2^*(\mf x)e^{-\mathrm{i}\Omega_2 \tau_2}\langle\hat{\psi}(\mf x) \rangle_{{\rho}_{\phi}}\ket{g_1g_2}\!\!\bra{g_1e_2}\Bigg).
    \end{align}
    Denoting $\hat{\rho}^{(l,k)} = \hat{U}_I^{(l)}\hat{\rho}_0\hat{U}_I^{(k)\dagger}$, we obtain the \tbb{term $\rho_{\text{d}}^{(1,1)}$ in Eq. \eqref{rho11}},
    \begin{align}
        \hat{\rho}_{\text{d}}^{(1,1)} =\lambda^2 \int \dd V\dd V'\Bigg(& \Lambda_1(\mf x)\Lambda_1^*(\mf x')e^{\mathrm{i}\Omega_1( \tau_1-\tau_1')}\langle\hat{\psi}(\mf x')\hat{\psi}^\dagger(\mf x) \rangle_{{\rho}_{\phi}}\ket{e_1g_2}\!\!\bra{e_1g_2}\\
        &\:\:\:\:\:+\Lambda_1(\mf x)\Lambda_2^*(\mf x')e^{\mathrm{i}(\Omega_1 \tau_1-\Omega_2 \tau_2')}\langle\hat{\psi}(\mf x')\hat{\psi}^\dagger(\mf x) \rangle_{{\rho}_{\phi}}\ket{e_1g_2}\!\!\bra{g_1e_2}\nonumber\\[8pt]
        &\:\:\:\:\:\:\:\:\:\:+\Lambda_2(\mf x)\Lambda_2^*(\mf x')e^{\mathrm{i}\Omega_2( \tau_2-\tau_2')}\langle\hat{\psi}(\mf x')\hat{\psi}^\dagger(\mf x) \rangle_{{\rho}_{\phi}}\ket{g_1e_2}\!\!\bra{g_1e_2}\nonumber\\
        &\:\:\:\:\:\:\:\:\:\:\:\:\:\:\:+\Lambda_2(\mf x)\Lambda_1^*(\mf x')e^{\mathrm{i}(\Omega_2 \tau_2-\Omega_1 \tau_1')}\langle\hat{\psi}(\mf x')\hat{\psi}^\dagger(\mf x) \rangle_{{\rho}_{\phi}}\ket{g_1e_2}\!\!\bra{e_1g_2}\Bigg).\nonumber
    \end{align}
    In order to compute the first of the remaining terms, notice that
    \begin{align}
        \nonumber \hat{h}_I(\mf x)\hat{h}_I(\mf x')\hat{\rho}_{\text{d}, 0} = &\lambda^2\Big(\Lambda_1(\mf x)e^{\mathrm{i}\Omega_1 \tau_1}\hat{\sigma}_1^+\hat{\psi}^\dagger(\mf x) + \Lambda^*_1(\mf x)e^{-\mathrm{i}\Omega_1 \tau_1}\hat{\sigma}_1^-\hat{\psi}(\mf x)+\Lambda_2(\mf x)e^{\mathrm{i}\Omega_2 \tau_2}\hat{\sigma}_2^+\hat{\psi}^\dagger(\mf x) + \Lambda^*_2(\mf x)e^{-\mathrm{i}\Omega_2 \tau_2}\hat{\sigma}_2^-\hat{\psi}(\mf x)\Big)\\
         &\:\:\:\:\:\:\:\:\:\:\:\:\:\:\:\:\:\:\:\:\:\:\:\:\:\:\:\:\:\:\:\:\:\:\:\:\:\:\:\:\:\:\:\:\:\:\:\:\:\:\:\:\:\:\:\:\:\:\:\:\:\:\:\:\:\:\:\:\:\:\:\:\:\:\:\:\:\:\:\:\:\:\:\:\:\:\:\times\Big(\Lambda_1(\mf x')e^{\mathrm{i}\Omega_1 \tau_1'}\hat{\sigma}_1^+\hat{\psi}^\dagger(\mf x') +\Lambda_2(\mf x')e^{\mathrm{i}\Omega_2 \tau_2'}\hat{\sigma}_2^+\hat{\psi}^\dagger(\mf x')\Big)\hat{\rho}_{\text{d}, 0}\nonumber\\
         &=\lambda^2\Bigg(\Lambda_1(\mf x)\Lambda_2(\mf x')e^{\mathrm{i}\Omega_1 \tau_1 + \mathrm{i}\Omega_2 \tau_2'}\hat{\psi}^\dagger(\mf x)\hat{\psi}^\dagger(\mf x')\ket{e_1e_2}\!\!\bra{g_1g_2}\nonumber\\
         &\:\:\:\:\:\:\:\:\:\:\:\:\:\:\:\:+\Lambda^*_1(\mf x)\Lambda_1(\mf x')e^{\mathrm{i}\Omega_1 (\tau_1'-\tau_1)}\hat{\psi}(\mf x)\hat{\psi}^\dagger(\mf x')\ket{g_1g_2}\!\!\bra{g_1g_2}\nonumber\\[8pt]
         &\:\:\:\:\:\:\:\:\:\:\:\:\:\:\:\:\:\:\:\:\:\:\:\:\:\:+\Lambda_2(\mf x)\Lambda_1(\mf x')e^{\mathrm{i}\Omega_2\tau_2 +\mathrm{i}\Omega_1 \tau_1'}\hat{\psi}^\dagger(\mf x)\hat{\psi}^\dagger(\mf x')\ket{e_1e_2}\!\!\bra{g_1g_2}\nonumber\\
         &\:\:\:\:\:\:\:\:\:\:\:\:\:\:\:\:\:\:\:\:\:\:\:\:\:\:\:\:\:\:\:\:\:\:\:\:+\Lambda^*_2(\mf x)\Lambda_2(\mf x')e^{\mathrm{i}\Omega_2 (\tau_2'-\tau_2)}\hat{\psi}(\mf x)\hat{\psi}^\dagger(\mf x')\ket{g_1g_2}\!\!\bra{g_1g_2}\Bigg)
    \end{align}
    and $\hat{\rho}_{\text{d}}^{(2,0)} = \hat{\rho}_{\text{d}}^{(0,2)\dagger}$ .
    
    We can then rearrange the final result, in order to write them in the $\{\ket{g_1g_2}, \ket{g_1e_2}, \ket{e_1g_2}, \ket{e_1e_2}\}$ basis. We define $\mathcal{M}$ according to Eq. \eqref{expReal} such that the $\hat{\rho}_{\text{d}}^{(2,0)}$ and $\hat{\rho}_{\text{d}}^{(0,2)}$ terms can be written as
    \begin{align}
        \hat{\rho}_{\text{d}}^{(2,0)} =&\:\: \mathcal{M}\ket{e_1e_2}\!\!\bra{g_1g_2}\nonumber\\& -\lambda^2 \int \dd V \dd V' \theta(t-t')\Big(\Lambda_2(\mf x)\Lambda^*_2(\mf x')e^{-\mathrm{i}\Omega_2 (\tau_2'-\tau_2)}\langle\hat{\psi}(\mf x')\hat{\psi}^\dagger(\mf x)\rangle_{\rho_{\phi}}\nonumber\\
         &\:\:\:\:\:\:\:\:\:\:\:\:\:\:\:\:\:\:\:\:\:\:\:\:\:\:\:\:\:\:\:\:\:\:\:\:\:\:\:\:\:\:\:\:\:\:\:\:\:\:\:+\Lambda_1(\mf x)\Lambda^*_1(\mf x')e^{-\mathrm{i}\Omega_1 (\tau_1'-\tau_1)}\langle\hat{\psi}(\mf x')\hat{\psi}^\dagger(\mf x)\rangle_{\rho_{\phi}}\Big)\ket{g_1g_2}\!\!\bra{g_1g_2},\\
        \hat{\rho}_{\text{d}}^{(0,2)} =&\:\:\mathcal{M}^* \ket{g_1g_2}\!\!\bra{e_1e_2} \nonumber\\
        &-\lambda^2 \int \dd V \dd V' \theta(t-t')\Big(\Lambda^*_2(\mf x)\Lambda_2(\mf x')e^{\mathrm{i}\Omega_2 (\tau_2'-\tau_2)}\langle\hat{\psi}(\mf x)\hat{\psi}^\dagger(\mf x')\rangle_{\rho_{\phi}}\nonumber\\
         &\:\:\:\:\:\:\:\:\:\:\:\:\:\:\:\:\:\:\:\:\:\:\:\:\:\:\:\:\:\:\:\:\:\:\:\:\:\:\:\:\:\:\:\:\:\:\:\:\:\:\:+\Lambda^*_1(\mf x)\Lambda_1(\mf x')e^{\mathrm{i}\Omega_1 (\tau_1'-\tau_1)}\langle\hat{\psi}(\mf x)\hat{\psi}^\dagger(\mf x')\rangle_{\rho_{\phi}}\Big)\ket{g_1g_2}\!\!\bra{g_1g_2}\nonumber\\
        =& \:\:\mathcal{M}^*\ket{g_1g_2}\!\!\bra{e_1e_2} \nonumber\\
        &-\lambda^2 \int \dd V \dd V' \theta(t'-t)\Big(\Lambda^*_2(\mf x')\Lambda_2(\mf x)e^{-\mathrm{i}\Omega_2 (\tau_2'-\tau_2)}\langle\hat{\psi}(\mf x')\hat{\psi}^\dagger(\mf x)\rangle_{\rho_{\phi}}\nonumber\\
         &\:\:\:\:\:\:\:\:\:\:\:\:\:\:\:\:\:\:\:\:\:\:\:\:\:\:\:\:\:\:\:\:\:\:\:\:\:\:\:\:\:\:\:\:\:\:\:\:\:\:\:+\Lambda^*_1(\mf x')\Lambda_1(\mf x)e^{-\mathrm{i}\Omega_1 (\tau_1'-\tau_1)}\langle\hat{\psi}(\mf x')\hat{\psi}^\dagger(\mf x)\rangle_{\rho_{\phi}}\Big)\ket{g_1g_2}\!\!\bra{g_1g_2}.
    \end{align}
    We further define $\mathcal{L}_{ii}$ and $\mathcal{E}_i$ according to Eq. \tbb{\eqref{antiexpgg}}.
    Using these definitions, we can write the evolved state in matrix form as
    \begin{equation}\label{mano}
        \hat{\rho}_{\text{d}} = \begin{pmatrix}
            1 - \L_{11}-\L_{22} & \mathrm{i}\mathcal{E}_2\color{black} & \mathrm{i}\mathcal{E}_1\color{black} & \mathcal{M}^*\\
             -\mathrm{i}\mathcal{E}_2^* & \L_{22} & \L_{21} & 0\\
            -\mathrm{i}\mathcal{E}_1^* & \L_{12} & \L_{11} & 0\\
            \mathcal{M} & 0 & 0 & 0 
        \end{pmatrix}.
    \end{equation}

\color{black}

\section{Computations regarding the \tb{excited-ground} interaction of the linear complex detector}\label{app:ge}
    Here we show the details of the computations regarding the general \tb{excited-ground} entanglement harvesting setup described in \tb{Subsection \ref{sub:antiharvestge}. Once again, notice that by considering a real spacetime smearing function and \mbox{$\hat{\psi}(\mf x) = \hat{\phi}(\mf x)$}, we can apply the results of this appendix to Subsection \ref{sub:UDWharvestge}.} Let us perform the same procedure as the one from Appendix \ref{app:gg}, but now considering the first detector to begin in the excited state $\ket{e_1}$. We then have that the initial density operator is
    \begin{equation}
        \hat{\rho}_0 = \ket{e_1}\!\!\bra{e_1}\otimes\ket{g_2}\!\!\bra{g_2}\otimes \hat{\rho}_{\phi} = \hat{\rho}_{\text{d}, 0}\otimes \hat{\rho}_{\phi}.
    \end{equation}
    Expanding the unitary time evolution operator to second order in the Dyson expansion as in Appendix \ref{app:gg}, we obtain the following expression for the final state of the detectors
    \begin{align}
        \hat{\rho}_{\text{d}} = \tr_\phi(\hat{\rho}) =&\hat{\rho}_{\text{d}, 0}-\text{i}\int \dd V\left(\ev{\hat{h}_I(\mf x)}_{\rho_{\phi}}\hat{\rho}_{\text{d}, 0}-\hat{\rho}_{\text{d}, 0}\ev{\hat{h}_I(\mf x)}_{\rho_{\phi}}\right)\nonumber\\&+\int \dd V\dd V'\left(\ev{\hat{h}_I(\mf x)\hat{\rho}_{\text{d}, 0}\hat{h}_I(\mf x')}_{\rho_{\phi}}-\theta(t-t')\left(\ev{\hat{h}_I(\mf x)\hat{h}_I(\mf x')}_{\rho_{\phi}}\hat{\rho}_{\text{d}, 0}+\hat{\rho}_{\text{d}, 0}\ev{\hat{h}_I(\mf x')\hat{h}_I(\mf x)}_{\rho_{\phi}}\right)\right),
    \end{align}
    where
    \begin{align}
        \hat{\rho}_{\text{d}}^{(1)} &= -\text{i}\lambda \int \dd V\Bigg( \Lambda_1^*(\mf x)e^{-\text{i}\Omega_1 \tau_1}\langle\hat{\psi}(\mf x) \rangle_{{\rho}_{\phi}}\ket{g_1g_2}\!\!\bra{e_1g_2}+\Lambda_2(\mf x)e^{\text{i}\Omega_2 \tau_2}\langle\hat{\psi}^\dagger(\mf x) \rangle_{{\rho}_{\phi}}\ket{e_1e_2}\!\!\bra{e_1g_2} \nonumber\\
        &\:\:\:\:\:\:\:\:\:\:\:\:\:\:\:\:\:\:\:\:\:\:\:\:\:\:\:\:\:\:\:\:-\Lambda_1(\mf x)e^{\text{i}\Omega_1 \tau_1}\langle\hat{\psi}^\dagger(\mf x) \rangle_{{\rho}_{\phi}}\ket{e_1g_2}\!\!\bra{g_1g_2}-\Lambda_2^*(\mf x)e^{-\text{i}\Omega_2 \tau_2}\langle\hat{\psi}(\mf x) \rangle_{{\rho}_{\phi}}\ket{e_1g_2}\!\!\bra{e_1e_2}\Bigg),\\
        \hat{\rho}_{\text{d}}^{(1,1)} &=\lambda^2 \int \dd V \dd V'\Bigg( \Lambda_1^*(\mf x)\Lambda_1(\mf x')e^{-\text{i}\Omega_1( \tau_1-\tau_1')}\langle\hat{\psi}^\dagger(\mf x')\hat{\psi}(\mf x) \rangle_{{\rho}_{\phi}}\ket{g_1g_2}\!\!\bra{g_1g_2}\nonumber\\
         &\:\:\:\:\:\:\:\:\:\:\:\:\:\:\:\:\:\:\:\:\:\:\:\:\:\:\:\:\:\:\:\:\:\:\:\:+\Lambda_1^*(\mf x)\Lambda_2^*(\mf x')e^{-\text{i}(\Omega_1\tau_1+\Omega_2 \tau_2')}\langle\hat{\psi}(\mf x')\hat{\psi}(\mf x) \rangle_{{\rho}_{\phi}}\ket{g_1g_2}\!\!\bra{e_1e_2}\nonumber\\[8pt]
        &\:\:\:\:\:\:\:\:\:\:\:\:\:\:\:\:\:\:\:\:\:\:\:\:\:\:\:\:\:\:\:\:\:\:\:\:\:\:\:\:+\Lambda_2(\mf x)\Lambda_2^*(\mf x')e^{\text{i}\Omega_2( \tau_2-\tau_2')}\langle\hat{\psi}(\mf x')\hat{\psi}^\dagger(\mf x) \rangle_{{\rho}_{\phi}}\ket{e_1e_2}\!\!\bra{e_1e_2}\nonumber\\
         &\:\:\:\:\:\:\:\:\:\:\:\:\:\:\:\:\:\:\:\:\:\:\:\:\:\:\:\:\:\:\:\:\:\:\:\:\:\:\:\:\:\:\:\:+\Lambda_2(\mf x)\Lambda_1(\mf x')e^{\text{i}(\Omega_2 \tau_2+\Omega_1 \tau_1')}\langle\hat{\psi}^\dagger(\mf x')\hat{\psi}^\dagger(\mf x) \rangle_{{\rho}_{\phi}}\ket{e_1e_2}\!\!\bra{g_1g_2}\Bigg),
    \end{align}
    and
    \begin{align}
         \hat{h}_I(\mf x)\hat{h}_I(\mf x')\hat{\rho}_{\text{d}, 0}
         &=\lambda^2\Bigg(\left(\Lambda_1^*(\mf x)\Lambda_2(\mf x')e^{\text{i}(-\Omega_1 \tau_1 + \Omega_2 \tau_2')}\hat{\psi}(\mf x)\hat{\psi}^\dagger(\mf x')+\Lambda_2(\mf x)\Lambda_1^*(\mf x')e^{\text{i}(\Omega_2 \tau_2 -\Omega_1 \tau_1')}\hat{\psi}^\dagger(\mf x)\hat{\psi}(\mf x')\right)\ket{g_1e_2}\!\!\bra{e_1g_2}\nonumber\\
         &\:\:\:\:\:\:\:\:\:\:\:\:+\left(\Lambda^*_2(\mf x)\Lambda_2(\mf x')e^{\text{i}\Omega_2 (\tau_2'-\tau_2)}\hat{\psi}(\mf x)\hat{\psi}^\dagger(\mf x')+\Lambda_1(\mf x)\Lambda_1^*(\mf x')e^{\text{i}\Omega_1 (\tau_1'-\tau_1)}\hat{\psi}^\dagger(\mf x)\hat{\psi}(\mf x')\right)\ket{e_1g_2}\!\!\bra{e_1g_2}\Bigg).
    \end{align}

    The evolved density operator in matrix form is then given by
    \begin{equation}\label{manoo}
        \hat{\rho}_{\text{d}} = \begin{pmatrix}
            {\tb{\mathcal{L}_{11}'}} & 0 & -\mathrm{i}\mathcal{E}_1 & \tb{\mathcal{L}_{12}'}\\
            0 & 0 & \mathcal{M}' & 0 \\
        \mathrm{i}\mathcal{E}_1^* & \mathcal{M}'^* & 1 - {\tb{\mathcal{L}_{11}'}} - \mathcal{L}_{22} & \mathrm{i}\mathcal{E}_2\\
            \tb{\mathcal{L}_{12}'}^* & 0 & -\mathrm{i}\mathcal{E}_2^* & \mathcal{L}_{22} 
        \end{pmatrix}.
    \end{equation}
    where we used the definitions for {$\tb{\mathcal{L}_{11}'}$,} $\tb{\mathcal{L}_{12}'}$ and $\mathcal{M}'$ from Eq. \eqref{Nijc}.
    
    \tb{In Eq. \eqref{manoo}, we confirmed the result that was highlighted in \cite{antiparticles}. Namely, we see that in order to change from the ground to the excited cases in the complex particle detector model, one can perform the changes $\Omega_i\longmapsto -\Omega_i$, $\hat{\psi}(\mf x)\longmapsto \hat{\psi}^\dagger(\mf x)$ and $\Lambda_i(\mf x) \longmapsto \Lambda_i^*(\mf x)$, where $i$ stands for the excited detector. Indeed, under these transformations, we notice that the matrices in Eqs. \eqref{manoo} and \eqref{mano} are related simply by reordering the basis elements associated to the first detector. This operation can equivalently be phrased as $\Omega_i\longmapsto -\Omega_i$, $\hat{\psi}_i(\mf x)\longmapsto \hat{\psi}_i^\dagger(\mf x)$. Writing it in this way then allows one to deduce the corresponding modifications for the fermionic model. Namely, in order to obtain the excited state from the ground state with the linearly coupled fermionic particle detector, one must perform the operations $\Omega_i\longmapsto -\Omega_i$, $\hat{\psi}(\mf x)\longmapsto \hat{\bar{\psi}}(\mf x)$ and $\Lambda_i(\mf x) \longmapsto \overline{\Lambda}_i(\mf x)$.}

    \color{black}
    \section{Entanglement Harvested by the linear complex scalar model in $(n+1)$-dimensional spacetime}\label{app:angles}
    
    In this appendix, we perform the computations of the $\mathcal{L}(\Omega)$ and $\mathcal{M}'$ terms of Section \ref{sec:antiHarvest} that lead to Eqs. \eqref{groundexcitedM} and \eqref{resultgroundexcited} in the case of $4$ dimensional spacetime. In this appendix we will perform the computations in $(n+1)$ dimensions, and later reduce them to $(3+1)$. Our final goal is to write both $\mathcal{L}(\Omega)$ and $\mathcal{M}'$ as integrals over one real variable, so we will assume that $n>1$ in this appendix.
    
    Both the $\mathcal{L}(\Omega)$ and $\mathcal{M}'$ terms involve $n$ dimensional integrals in the momentum variable $\bm p$. Given that most of the integrands depend only on $|\bm p|$, we pick spherical coordinates to solve the integrals. In spherical coordinates, it is then possible to write the integration measure $\dd^n \bm p $ as
    \begin{equation}\label{eq:c1}
        \dd^n \bm p = |\bm p|^{n-1}\dd |\bm p| \dd\Omega_{n-1} = |\bm p|^{n-1}(\sin\theta)^{n-2}\dd |\bm p| \dd\theta \dd \Omega_{n-2},
    \end{equation}
    where $\dd\Omega_{m}$ denotes the volume element of the $m$ dimensional unit sphere, $S^m$ and $\theta$ is the polar angle in $n$ dimensions. We have
    \begin{equation}
        \int \dd\Omega_{m} = \frac{2\pi^\frac{m+1}{2}}{\Gamma\left(\frac{m+1}{2}\right)},
    \end{equation}
    where $\Gamma(m)$ is the Gamma function. This result allows us to solve for the $\mathcal{L}(\Omega)$ term, under the assumption that $\tilde{F}(\bm p)$ only depends on $|\bm p|$ (real symmetric smearing function). We obtain:
    \begin{equation}
        \mathcal{L}(\Omega) = \frac{\lambda^2}{(2\pi)^n}\!\int \frac{\dd^n \bm p}{2 \omega_{\bm p}}|\tilde{F}(\bm p)|^2 |\tilde{\chi}( \omega_{\bm p}-\Omega)|^2
        = {\color{black}\frac{2\lambda^2}{({4}\pi)^\frac{n}{2} \Gamma(\frac{n}{2})}} \int_0^\infty \frac{\dd |\bm  p|}{2 \omega_{\bm p}} |\bm p|^{n-1} \:|\tilde{F}(|\bm p|)|^2 |\tilde{\chi}( \omega_{\bm p}-\Omega)|^2.
    \end{equation}
    
    However, even if $F(\bm x)$ is spherically symmetric, the $\mathcal{M}'$ term depends on the inner product $\bm p \cdot \bm L$. {\color{black} By choosing coordinates for the momentum integral} so that the $z$-axis to be aligned with $\bm L$, we can write $\bm p \cdot \bm L = |\bm p||\bm L|\cos\theta$, so that the $\mathcal{M}'$ term reads
    \begin{align}\label{eq:c4}
        \mathcal{M}'=&\frac{\lambda^2}{(2\pi)^n} \!\int \frac{\dd^n \bm p}{2 \omega_{\bm p}} |\tilde{F}(\bm p)|^2 e^{\text{i} \bm p \cdot \bm L}\!\tbb{\Big(Q(\Omega-\omega_{\bm p})+Q(-\omega_{\bm p}-\Omega)\Big)},\\
        =&\frac{\lambda^2}{(2\pi)^n} \int_0^\infty \frac{\dd |\bm  p|}{2 \omega_{\bm p}} |\bm p|^2 \:|\tilde{F}(|\bm p|)|^2 \tbb{\Big(Q(\Omega-\omega_{\bm p})+Q(-\omega_{\bm p}-\Omega)\Big)} \int \dd\Omega_{n-1} e^{\text{i}|\bm p||\bm L| \cos\theta}.\nonumber
    \end{align}
   {\color{black}  Let $\text{d}\Omega_n$ be the area element of the hypersphere of dimension $n>1$. Using angular coordinates $\{\theta_1,...,\theta_n\}$ such that $\theta_1$ ranges from $0$ to $2\pi$ and $\theta_i$ ranges from $0$ to $\pi$ for $i  = 2,...,n$, the area element can be written as
   \begin{equation}
       \dd \Omega_n = \prod_{i=1}^n (\sin\theta_{i})^{i-1} \dd \theta_i.
   \end{equation}
   Then, notice that the area element of the hypersphere of dimension $n-1$ is the area element of the hypersphere of dimension $n-2$ times $\dd\theta\,(\sin\theta)^{n-2}$. In fact,
   \begin{align}
	    \dd\Omega_{n-1} = \dd\theta_{n-1} (\sin\theta_{n-1})^{n-2}\prod_{i=1}^{n-2}\dd\theta_i\,(\sin\theta_i)^{i-1}= \dd\Omega_{n-2}\dd\theta_{n-1}\,(\sin\theta_{n-1})^{n-2}.
	\end{align}
    Note that
	\begin{align}
	    \int \dd\Omega_{n-1} &= \frac{2 \pi ^{n/2}}{\Gamma \left(\frac{n}{2}\right)}\,,\\
	    \int_0^\pi\dd\theta (\sin\theta)^{n-2}  &= \frac{\sqrt{\pi } \Gamma \left(\frac{n-1}{2}\right)}{\Gamma \left(\frac{n}{2}\right)}\,,
	\end{align}
    hence we can write
    \begin{align}
        \int \dd\Omega_{n-1} &= \frac{\int \dd\Omega_{n-1}}{\int_0^\pi\dd\theta (\sin\theta)^{n-2} } 
        = \frac{2 \pi ^{\frac{n-1}{2}}}{\Gamma \left(\frac{n-1}{2}\right)}\,.
    \end{align}}
	
	Considering all this, the integral over the angular variables can be solved analytically. Notice that for our expressions (e.g. \eqref{eq:c1} and \eqref{eq:c4}) we notated $\theta \equiv \theta_{n-1}$. We hence obtain
    \begin{align}
        \int \dd\Omega_{n-1} e^{\text{i}|\bm p||\bm L| \cos\theta} &= \int \dd\Omega_{n-2} \int_0^\pi \dd\theta\,(\sin\theta)^{n-2} e^{\text{i}|\bm p||\bm L| \cos\theta} = \frac{2\pi^\frac{n-1}{2}}{\Gamma\left(\frac{n-1}{2}\right)}\int_0^\pi \dd\theta\,(\sin\theta)^{n-2} e^{\text{i}|\bm p||\bm L| \cos\theta} \nonumber\\
        &= \frac{2\pi^\frac{n-1}{2}}{\Gamma\left(\frac{n-1}{2}\right)}\int_{-1}^1 \dd u\,(1-u^2)^{\frac{n-3}{2}} e^{\text{i}|\bm p||\bm L|u} = 2 \pi^{\frac{n}{2}}{}_0F_1\left(\frac{n}{2},-\frac{|\bm p|^2|\bm L|^2}{4}\right),
    \end{align}
    where ${}_0F_1(x)$ denotes the hypergeometric function. We then obtain an expression for $\mathcal{M}$ in terms of a single integral,
    \begin{equation}
        \mathcal{M}'={\color{black}\frac{2\lambda^2}{(4\pi)^\frac{n}{2}}} \int_0^\infty \frac{\dd |\bm  p|}{2 \omega_{\bm p}} |\bm p|^2 \:|\tilde{F}(|\bm p|)|^2 \:{}_0F_1\left(\frac{n}{2},-\frac{|\bm p|^2|\bm L|^2}{4}\right)\tbb{\Big(Q(\Omega-\omega_{\bm p})+Q(-\omega_{\bm p}-\Omega)\Big)}.
    \end{equation}
    In the particular case of $n = 3$, the hypergeometric function reduces to
    \begin{equation}
        {}_0F_1\left(\frac{3}{2},-\frac{|\bm p|^2|\bm L|^2}{4}\right) = \frac{2}{\sqrt{\pi}} \frac{\sin(|\bm p||\bm L|)}{|\bm p||\bm L|} = \frac{2}{\sqrt{\pi}}\text{sinc}(|\bm p||\bm L|).
    \end{equation}

    \twocolumngrid

\bibliography{references}
    
\end{document}